\documentclass[twocolumn]{aastex63}

\usepackage{hyperref}
\usepackage{xcolor}
\usepackage{url}
\usepackage{amssymb}
\usepackage{amsmath}
\hypersetup{
    colorlinks,
    linkcolor={red!50!black},
    citecolor={blue!50!black},
    urlcolor={black} 
}
\usepackage{mathrsfs}
\usepackage{nicefrac}
\usepackage{soul}

\newcommand{\oiii}{[\ion{O}{3}]}
\newcommand{\nii}{[\ion{N}{2}]}
\newcommand{\sii}{[\ion{S}{2}]}
\newcommand{\ha}{H$\rm\alpha$}
\newcommand{\hb}{H$\rm\beta$}
\newcommand{\kms}{km $\rm s^{-1}$}

\graphicspath{{./}{figures/}}

\received{August 7, 2020}
\revised{October 12, 2020}
\accepted{October 20, 2020}
\submitjournal{ApJ}
\shorttitle{Fast Outflows in Hot DOGs}
\shortauthors{Finnerty et al.}

\begin{document}

\title{Fast Outflows in Hot Dust-Obscured Galaxies Detected with Keck/NIRES}

\correspondingauthor{Luke Finnerty}
\email{lfinnerty@astro.ucla.edu}

\author[0000-0002-1392-0768]{Luke Finnerty}
\affiliation{Division of Physics, Math, and Astronomy, California Institute of Technology, \\
1200 E California Blvd, Pasadena, California, 91125, USA}

\author{Kirsten Larson}
\affiliation{Division of Physics, Math, and Astronomy, California Institute of Technology, \\
1200 E California Blvd, Pasadena, California, 91125, USA}

\author{B.T. Soifer}
\affiliation{Division of Physics, Math, and Astronomy, California Institute of Technology, \\
1200 E California Blvd, Pasadena, California, 91125, USA}

\author{Lee Armus}
\affiliation{IPAC, California Institute of Technology, 1200 E California Blvd,  Pasadena, CA 91125, USA}

\author{Keith Matthews}
\affiliation{Division of Physics, Math, and Astronomy, California Institute of Technology, \\
1200 E California Blvd, Pasadena, California, 91125, USA}

\author[0000-0003-1470-5901]{Hyunsung D. Jun}
\affiliation{School of Physics, Korea Institute for Advanced Study, 85 Hoegiro, Dongdaemun-gu, Seoul 02455, Republic of Korea}

\author[0000-0002-0786-7307]{Dae-Sik Moon}
\affiliation{Dept. of Astronomy and Astrophysics, University of Toronto, 50 St. George Street, Toronto, ON M5S 3H4, Canada}

\author{Jason Melbourne}
\affiliation{Division of Physics, Math, and Astronomy, California Institute of Technology, \\
1200 E California Blvd, Pasadena, California, 91125, USA}

\author{Percy Gomez}
\affiliation{W.M. Keck Observatory, Kameula, HI, USA}

\author[0000-0002-9390-9672]{Chao-Wei Tsai}
\affiliation{National Astronomical Observatories, Chinese Academy of Sciences, 20A Datun Road, Chaoyang District, Beijing 100012, People's Republic of China}

\author{Tanio Diaz-Santos}
\affiliation{Núcleo de Astronomía de la Facultad de Ingeniería, Universidad Diego Portales, Av. Ejército Libertador 441, Santiago, Chile}
\affiliation{Chinese Academy of Sciences South America Center for Astronomy (CASSACA), National Astronomical Observatories, CAS, Beijing 100101, China}
\affiliation{Institute of Astrophysics, Foundation for Research and Technology—Hellas (FORTH), Heraklion, GR-70013, Greece}

\author{Peter Eisenhardt}
\affiliation{Jet Propulsion Laboratory, California Institute of Technology, 4800 Oak Grove Drive, Pasadena, CA 91109, USA}

\author{Michael Cushing}
\affiliation{Department of Physics and Astronomy, University of Toledo, 2801 West Bancroft Street, Toledo, OH 43606, USA}

\begin{abstract}
We present rest-frame optical spectroscopic observations of 24 Hot Dust-Obscured Galaxies (Hot DOGs) at redshifts 1.7--4.6 with KECK/NIRES. Our targets are selected based on their extreme red colors to be the highest luminosity sources from the WISE infrared survey. In 20 sources with well-detected emission we fit the key \oiii, \hb, \ha, \nii, and \sii\ diagnostic lines to constrain physical conditions. Of the 17 targets with a clear detection of the \oiii$\rm \lambda$5007 \r A emission line, 15 display broad blueshifted and asymmetric line profiles, with widths ranging from 1000 to 8000 \kms\ and blueshifts up to 3000 \kms. These kinematics provide strong evidence for the presence of massive ionized outflows of up to $8000\ \rm M_\odot\ yr^{-1}$, with a median of $150\ \rm M_\odot\ yr^{-1}$. As many as eight sources show optical emission line ratios consistent with vigorous star formation. Balmer line star-formation rates, uncorrected for reddening, range from 30--1300 $\rm M_\odot\ yr^{-1}$, with a median of $50\ \rm M_\odot\ yr^{-1}$. Estimates of the SFR from SED fitting of mid and far-infrared photometry suggest significantly higher values. We estimate the central black hole masses to be of order $10^{8-10}\rm\ M_\odot$, assuming the present-day $\rm M_{BH}-\sigma_*$ relation. The bolometric luminosities and the estimated masses of the central black holes of these galaxies suggest that many of the AGN-dominated Hot DOGs are accreting at or above their Eddington limit. The combination of ongoing star formation, massive outflows, and high Eddington ratios suggest Hot DOGs are a transitional phase in galaxy evolution.
\end{abstract}


\section{Introduction}
Hot, Dust-Obscured Galaxies \citep[Hot DOGs,][]{Eisenhardt_2012, Wu_2012} are a population selected through the ``W1W2-dropout" criteria from WISE mission photometry \citep{Wright_2010}.  These objects are well detected in the WISE 12 $\mu$m and 22 $\mu$m bands (W3 and W4), but are detected poorly or not at all in the 3.4 $\mu$m and 4.6 $\mu$m bands (W1 and W2) \citep{Eisenhardt_2012}, indicating an extremely red Spectral Energy Distribution (SED). Specifically, the selection criteria require (in Vega magnitude units) W1 $>$ 17.4 and [W2$-$W4] $>$ 8.2.  Hot DOGs usually have redshifts in the range of 2--3, and are relatively rare with approximately 1000 detected in the WISE All-Sky data release \citep{Cutri_2012}, although it has been suggested that they may be as common as equally luminous unobscured quasars \citep{Assef_2015}. This suggests these objects may be an important, short-lived phase of galaxy evolution near the peak of both star formation and supermassive black hole growth \citep{Wu_2012, bridge_2013, Madau_2014}.

Subsequent follow-up observations have emphasized the extreme nature of Hot DOGs. Hot DOGs have luminosities above $10^{13}\ \rm L_\odot$, making them among the most luminous galaxies in the Universe \citep{Wu_2012, Assef_2015, Tsai_2015}. As the name suggests, Hot DOGs have warmer dust than normal Dust-Obscured Galaxies (DOGs), on the order of 100 K, leading to significant differences in the SEDs of Hot DOGs compared with submilimeter galaxies or normal DOGs \citep{Wu_2012, Melbourne_2012}.  The combination of warm dust and AGN activity observed in Hot DOGs led to the suggestion that Hot DOGs are obscured AGN with significant ongoing star formation in the host galaxy \citep{Eisenhardt_2012}. 

Previous observations of Hot DOGs have been focused on rest-frame infrared wavelengths, probing the hot dust emission that dominates the SED \citep[e.g.][]{Eisenhardt_2012, Tsai_2015}.  Additional work by \citet{Assef_2015} extended the SEDs to rest-frame optical/near-infrared ($<2\mu$m). Subsequent observations of the H$\alpha$ emission in five Hot DOGs detected broad emission in all targets, with Eddington ratios close to unity \citep{Wu_2018}. Full rest-frame optical spectroscopy of 12 targets was first published in \citet{Jun_2020}, who found ionized outflows in optical emission lines showing extreme kinematics, with typical \oiii$\lambda$5007\r A blueshifts of approximately 1100 \kms\ and full widths at half maximum (FWHMs) of approximately 2600 \kms.

Here, we present the results of Keck/NIRES observations covering the 0.95--2.4 $\mu$m  spectral range of 24 Hot DOGs (rest-frame optical/UV), 20 of which have not been previously reported. This work brings the total number of Hot DOGs with published rest-frame optical/UV spectra to 32. Section \ref{sec:obs} summarizes the observations obtained, data reduction procedures, and the total infrared luminosity estimates. Our spectral analysis procedures, including redshift determination and line profile fitting, are described in Section \ref{sec:extract}. Section \ref{sec:results} discusses the origin of the radiation source in Hot DOGs and the kinematics of the profile fits, followed by a discussion of the implied outflow, star formation, and black hole properties in Section \ref{sec:disc}. We summarize our results in Section \ref{sec:conc}. For all calculations, we assume a flat $\Lambda$CDM cosmology with $\Omega_M$ = 0.3, $\Omega_\Lambda$ = 0.7, and $H_0$ = 70 \kms\ $\rm Mpc^{-1}$.

\section{Observation and Data Reduction}\label{sec:obs}
\begin{figure*}
    \centering
    \noindent\includegraphics[width=40pc]{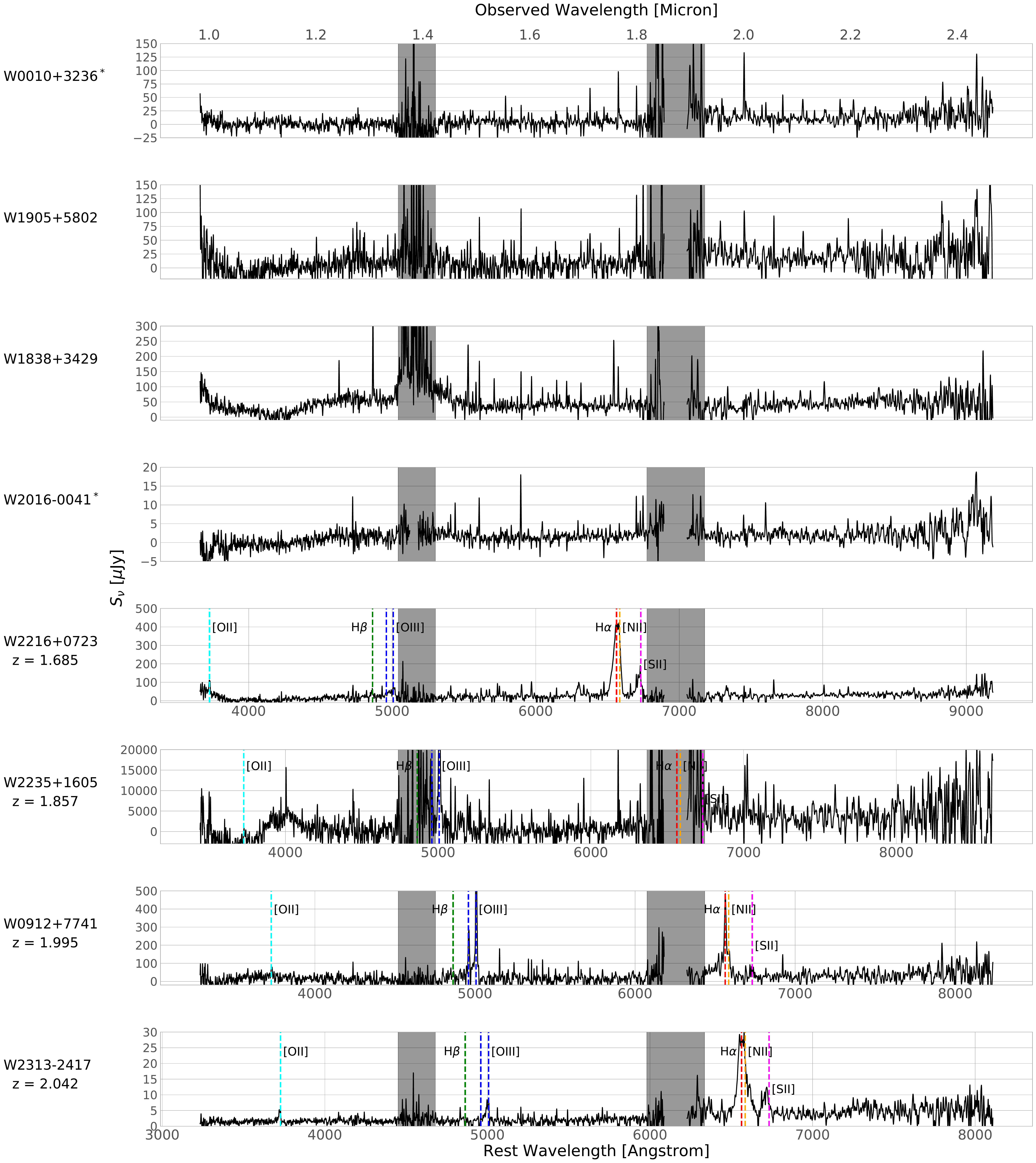}
    \caption{Full flux-calibrated NIRES spectra (in $\mu$Jy), with significant emission lines marked at their expected location based on the target redshift. Targets are sorted by systemic redshift.  Regions of high telluric absorption are shaded grey, and the spectrum has been convolved with a two pixel Gaussian kernel for clarity.  Figure continues on the next page. Objects marked $^*$ may have unreliable absolute flux calibration due to a lack of prior photometry or reference objects in the NIRES slit image.}
    \label{halfdogs1}
\end{figure*}

\begin{deluxetable*}{cccccccccccc}
\tablewidth{0pt}
\tabletypesize{\scriptsize}
\tablecaption{Observation and Target Properties}
\tablehead{ & \colhead{RA} & \colhead{Dec} & \colhead{$z$} & \colhead{$L_{bol}$} & \colhead{$L_{3.4-160}$} & \colhead{$L_{IR}$} &  \colhead{W2} & \colhead{W4} & \colhead{160$\mu$m} & Obs. Date & \colhead{$t_{int}$}  \\ & & & & $10^{13} L_\odot$ & $10^{13} L_\odot$ & $10^{13} L_\odot$ & $\mu$Jy & mJy & mJy & UT DD/MM/YY & min }
\startdata
W0010+3236 & 00:10:14.08 & +32:36:17.1 & - & - & - & - & 81$\pm$10 & 17$\pm$1 & - & 08/09/19 & 30 \\
W0116-0505 & 01:16:01.41 & -05:05:04.1 & 3.191 & 12.3 & 11.7 & 18.8 & 91$\pm$12 & 12$\pm$1 & 93$\pm$6$^b$ & 23/10/18 & 55 \\
W0255+3345 & 02:55:34.89 & +33:45:57.7 & 2.668 & 10.5 & 9.9 & 11.4 & 36$\pm$10 & 17$\pm$1 & 73$\pm$7$^b$ & 24/10/18 & 75 \\ 
W0220+0137 & 02:20:52.13 & +01:37:11.4 & 3.138 & 7.5 & 6.4 & 21.6 & 40$\pm$9 & 12$\pm$1 & 120$\pm$6$^b$ & 30/08/20 & 55  \\
W0338+1941 & 03:38:51.33 & +19:41:28.5 & 2.131 & 3.0 & 3.0 & 6.4 & 37$\pm$3 & 11$\pm$1 & 48$\pm$9$^e$ & 31/08/20 & 60 \\
W0410-0913 & 04:10:10.61 & -09:13:05.2 & 3.610 & 17.6 & 13.8 & 36.7 & 35$\pm$10 & 14$\pm$1 & 108$\pm$13$^c$ & 23/10/18 & 50 \\
W0514-1217 & 05:14:42.63 & -12:17:24.6 & 2.235 & 11.4 & 10.6 & 14.7 & 127$\pm$11 & 32$\pm$1 & 165$\pm$20$^d$ & 22/10/18 & 55 \\ 
W0831+0140 & 08:31:53.25 & +01:40:10.7 & 3.915 & 18.9 & 13.6 & 33.6 &  63$\pm$11 & 11$\pm$1 &$<60^b$ & 17/03/19 & 100 \\
W0859+4823 & 08:59:29.93 & +48:23:02.0 & 3.256 & 10.7 & 9.2 & 15.1 & 42$\pm$9 & 13$\pm$1 & 34$\pm$11$^c$ & 11/10/19 & 70 \\
W0912+7741 & 09:12:47.16 & +77:41:58.2 & 1.995 & 2.0 & 1.9 & 1.8 & 26$\pm$8 & 9$\pm$1 & 17$\pm$12$^d$ & 16/03/19 & 60 \\
W1322-0328 & 13:22:32.57 & -03:28:42.2 & 3.043 & 10.3 & 9.2 & 13.8 & 65$\pm$11 & 12$\pm$1 & 64$\pm$7$^b$ & 17/03/19 & 60 \\
W1719+0446 & 17:19:46.63 & +04:46:35.2 & 2.551 & 2.8 & 2.0 & 7.7 & 110$\pm$10 & 15$\pm$1 & 43$\pm$8$^d$ & 30/08/20 & 60  \\
W1724+3455 & 17:24:01.35 & +34:55:58.0 & 2.366 & 4.6 & 4.1 & 6.1 & 16$\pm$7 & 13$\pm$1 & 56$\pm$6$^e$ & 26/06/18 & 50 \\
W1801+1543 & 18:01:25.67 & +15:43:15.8 & 2.329 & 8.9 & 8.0 & 10.5 & 140$\pm$3  & 25$\pm$1 & 98$\pm$7$^{e}$ & 31/08/20 & 65 \\
W1835+4355 & 18:35:33.71 & +43:55:49.0 & 2.302 & 8.9 & 7.8 & 9.7 & 143$\pm$7 & 29$\pm$1 & 101$\pm$13$^c$ & 24/06/18 & 80 \\
W1838+3429 & 18:38:09.15 & +34:29:25.8 & - & - & - & - & 31$\pm$7 & 9$\pm$1 & 38$\pm$7$^b$ & 08/09/19 & 65 \\
W1905+5802 & 19:05:00.07 & +58:02:56.8 & - & - & - & - & 66$\pm$5 & 16$\pm$1 & - & 09/09/19 & 55 \\
W2216+0723 & 22:16:19.09 & +07:23:53.3 & 1.685 & 1.3 & 0.8 & 5.6 & 100$\pm$12 & 15$\pm$1 & 131$\pm$9$^c$ & 22/10/18 & 50 \\ 
W2016-0041 & 20:16:50.30 & -00:41:09.0 & - & - & - & - & 59$\pm$11 & 9$\pm$1 & - & 30/08/20 & 85 \\
W2235+1605 & 22:35:43.66 & +16:05:10.7 & 1.857$^*$ & 5.9 & 5.7 & 7.3 & 111$\pm$11 & 26$\pm$1 & 112$\pm$7$^e$ & 08/09/19 & 30 \\
W2238+2653 & 22:38:10.20 & +26:53:19.7 & 2.397 & 9.1 & 7.8 & 15.3 & 64$\pm$9 & 18$\pm$1 & 142$\pm$12$^c$ & 24/06/18 & 75 \\
W2246-0526 & 22:46:07.56 & -05:26:34.9 & $4.602^{*,a}$ & 32.7 & 28.8 & 65.9 & 38$\pm$13 & 16$\pm$2 & 125$\pm$12$^c$ & 23/10/18 & 75 \\
W2305-0039 & 23:05:25.88 & -00:39:25.7 & 3.108 & 18.1 & 16.5 & 21.2 & 67$\pm$11 & 25$\pm$1 & 128$\pm$13$^c$ & 10/11/19 & 75 \\
W2313-2417 & 23:13:01.56 & -24:17:56.8 & 2.042 & 3.8 & 3.4 & 6.0 & 60$\pm$3 & 14$\pm$2 & 91$\pm$6$^{e}$ & 31/08/20 & 95 \\
\enddata
\tablecomments{Coordinates and fluxes in W2 and W4 are taken from the AllWISE source catalog \citep{Cutri_2012}, values marked (a) are from \citet{Tsai_2018}, (b) are from \citet{Tsai_2015}, (c) are from \citet{Fan__2016}, (d) are from \citet{Farrah_2017}, and (e) are from Tsai et al. (in preparation). Redshifts are from cross-correlating the NIRES spectra with an SDSS linelist unless noted. * indicates a redshift determined from a single line in the NIRES spectrum. Bolometric luminosities are estimated using the approach from \citet{Tsai_2015} to fit MIR and FIR photometry from \citet{Wu_2012}, \citet{Tsai_2015}, \citet{Fan__2016}, \citet{Farrah_2017}, and Tsai et al. (in preparation) to obtain a conservative lower limit. $L_{3.4-160}$ uses the same approach but only the photometry from 3.4-160$\mu$m.  $L_{IR}$ estimates the total infrared luminosity from the by fitting the SEDs from \citet{chary_2001} to the available photometry with a linear scale factor. W2 and W4 fluxes are from the AllWISE source catalog \citep{Cutri_2012} at 4.6$\mu$m and 22$\mu$m respectively, and were converted from WISE magnitude units using the zeropoints from \citet{Jarrett_2011}.  160$\mu$m photometry is from Herschel. Note that W0010+3236, W0116-0505, and W0514-1217 do not satisfy the W1W2-dropout criteria, but are included due to the WISE colors and similar rest-frame optical spectra.}
\label{obsprops}
\end{deluxetable*}
Among other targets, a total of 24 Hot DOGs were observed over 12 nights using Keck/NIRES. NIRES is a cross-dispersed echellette spectrograph offering full simultaneous coverage of the $Y$, $J$, $H$, and $K$-bands over five echelle spectral orders at an average spectral resolution of 2700 with a $0\farcs55\times18\arcsec$ slit on the 10-m Keck 2 telescope \citep[Moon et al. in preparation, see also][]{nires}. The broad wavelength coverage makes NIRES ideal for targets with unknown or poorly-constrained redshifts, as was the case for several of the Hot DOGs. Table \ref{obsprops} contains target and observation parameters for the observed objects. Full 1D extracted spectra for each target are presented in Figure \ref{halfdogs1}, with prominent emission lines marked and regions of high telluric absorption shaded.

Observations were taken with five-minute individual exposures in ABAB or ABACA dither patterns. ABAB alternated between locations towards the top and bottom of the slit, while ABACA alternates between center, top, and bottom positions. The average total integration time was 65 minutes. Integration times varied as a result of weather conditions and scheduling considerations for other objects in the observing program. Observations of W0010+3236 and W2235+1605 were cut short when no features were apparent in the raw NIRES spectra in order to maximize time spent on other targets. For some sources (e.g. W2246-0526 and W0410-0913 in Figure \ref{halfdogs1}), the long individual exposures resulted in relatively poor sky subtraction, particularly in the $H$-band. Flat fields were obtained using the dome flat-fielding lamps.

Spectra were reduced using SpexTool \citep{spex} updated for NIRES. After flat-fielding, SpexTool performs a 2D wavelength calibration using approximately 200 sky lines computed with \citet{Lord_1992}. The total error in the wavelength calibration, including effects from instrument flexure, was on the order of 0.5 pixels. The calibration errors are significantly smaller than the 2.7 pixel resolution element of NIRES, and correspond to a velocity of 20 \kms\ at the average instrument resolution. For most line profile fits, the reported error is significantly larger than the calibration uncertainty, indicating the wavelength calibration is not the dominant source of error. Sky and dark subtraction used the exposure dithering. Seeing was typically greater than the 0\farcs55 NIRES slit (in the range of approximately 0\farcs4 -- 1\arcsec, averaging 0\farcs6 -- 0\farcs7), and all targets were effectively point-like in the NIRES trace. Trace centers were identified manually and extracted with the optimal extraction algorithm as described in \citet{spex}. For telluric correction, observations of an A0 standard star at similar airmass were taken immediately before or after each science observation and divided from the stacked 1D target spectrum. While this usually provides a high-quality telluric correction and relative flux calibration, detector persistence issues affecting observations on 8 and 9 September 2019 led to significant artifacts in the $y$ and $J$ band continua for observations from those nights (blueward of 1.35 $\mu$m in the NIRES spectrum). This is most clearly seen in W2235+1605, and to a lesser extend in W1838+3429. In all targets, poor corrections are obtained in the strong telluric absorption bands near 1.4 $\mu$m and 1.85 $\mu$m (see shaded regions in Figure \ref{halfdogs1}). In W2235+1605, these telluric features prevent clear identification and fitting of emission features, and the quality of the \oiii\ fit in W0255+3345 is impacted by telluric features. Other lines are well-separated from poorly-corrected absorption features and fitting does not appear to be significantly impacted by tellurics. In cases of poor sky subtraction, residual sky features are manually clipped during the emission line fitting process and do not significantly impact the final fit.

Absolute flux calibration was performed using the $K'$-band photometry from \citet{Assef_2015} when available. Targets without such photometry were flux-calibrated by comparison with 2MASS \citep{2MASS} sources in the NIRES $K'$-band slit-viewing camera, with a $1.8\arcmin\times1.8\arcmin$ field of view. By comparing different 2MASS calibration sources for the same objects, we estimate the uncertainty in the absolute flux calibration is 20--30 percent. No prior photometry or suitable 2MASS sources were available for W0010+3236, W0831+0140, W2016-0041, W2238+2653, and W2305-0039, all of which were observed under non-photometric conditions. We therefore omit these source from portions of the analysis which depend on the absolute flux, and note that the flux calibrations in Figure \ref{halfdogs1} are approximate for these objects. 

For targets with both \ha\ and \hb\ detections we report the Balmer decrements and implied $A_V$ values in Table \ref{ratioprops}. This provides an estimate for the global impact of reddening and extinction in those objects, which appears to be substantial. We use these extinction estimates to correct \ha\ luminosities in Section \ref{sub:sfrs}. Since it is not clear whether the average extinction, which may be dominated by dust in the main body of the galaxy, should be applied to outflowing gas, Table \ref{outprops} presents the derived outflow properties (outflow mass, mass outflow rate, energy/momentum fluxes) based on the uncorrected \oiii\ luminosity. The true values may be substantially larger, depending on the extinction of the outflow.

The total luminosity $\rm L_{bol}$ in Table \ref{obsprops} is estimated by adopting the technique outlined by \citet{Tsai_2015}, making a power-law interpolation between the mid/far infrared photometry from \citet{Wu_2012, Tsai_2015, Fan__2016, Farrah_2017}; and Tsai et al. (in prep.) for all targets and integrating with bounds extended 20 percent beyond the photometry.  This power-law based interpolation provides conservative estimates of the bolometric and total infrared luminosity compared with the torus+dust SEDs fit from \citet{Fan__2016} \citep{Tsai_2015}.  This approach is preferred over the use of a scaling relation based on an AGN model as described in \citet{Assef_2015} because it does not require a reddening correction based on other AGN templates.  The underestimation of the bolometric luminosities in Table \ref{obsprops} implies the estimates of the Eddington ratio in Table \ref{bhprops} are lower limits.  We also calculate the luminosity based on observed-frame 3.4--160 $\mu$m photometry, $\rm L_{3.4-160}$, for all targets to enable a more clear comparison of the luminosities, as some targets lack FIR photometry. We also estimate the total 8--1000 $\mu$m infrared luminosity, $\rm L_{IR}$, based on the  \citet{chary_2001} SED grid, with a linear scale factor applied to the SED in order to better match the extreme Hot DOG luminosities. The quality of SED fits to the Hot DOG photometry was poor due to the lack of a hot dust component in the model SEDs, and we therefore prefer the lower-limit $\rm L_{bol}$ and $\rm L_{3.4-160}$ estimates using the \citet{Tsai_2015} technique in subsequent analysis. The luminosities from the SED fits should be seen as a  first-order comparison to the luminosities of local ULIRGS. We also note that the use of a power-law interpolation to estimate luminosity may be inaccurate if the emission is significantly anisotropic \citep[e.g.][]{Richards_2006}, but recent work has found emission redward of 15 $\mu$m is highly isotropic \citep{almeida_2017}. 

W0514-1217 is not a Hot DOG according to the W1W2$-$dropout criterion due to its bright W1 flux, and was instead selected based on the [W2$-$W3] color criteria of \citet{bridge_2013}. We include W0514-1217 in subsequent analysis based on the similarity of the rest-frame optical spectrum to the targets which do satisfy the W1W2-dropout conditions. W0010+3236 and W0116-0505 also fail the W1 criteria, but by a much smaller margin than W0514-1217.
\begin{figure*}
    \centering
    \noindent\includegraphics[width=40pc]{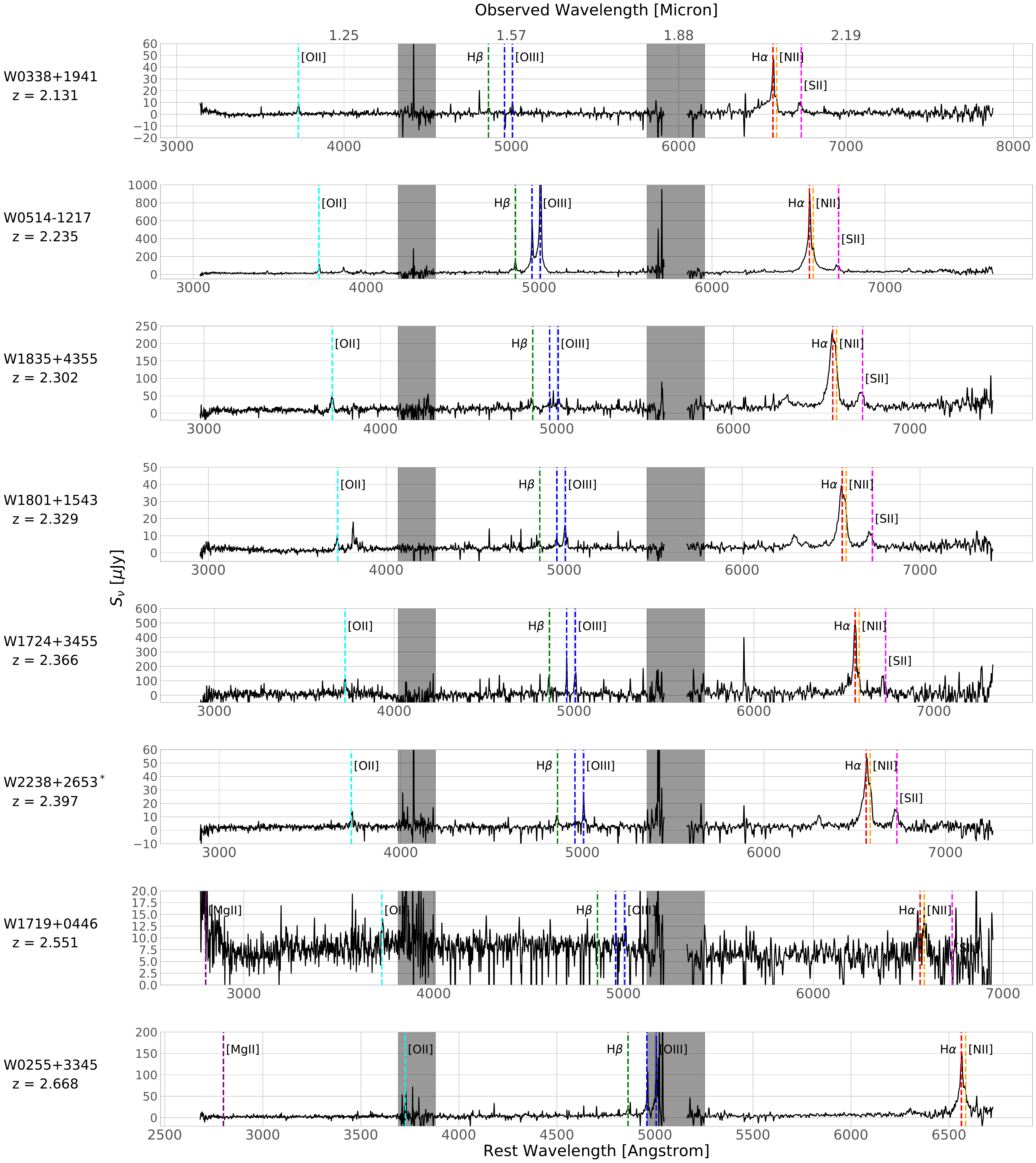}
    \label{halfdogs2}
\end{figure*}
\begin{figure*}
    \centering
    \noindent\includegraphics[width=40pc]{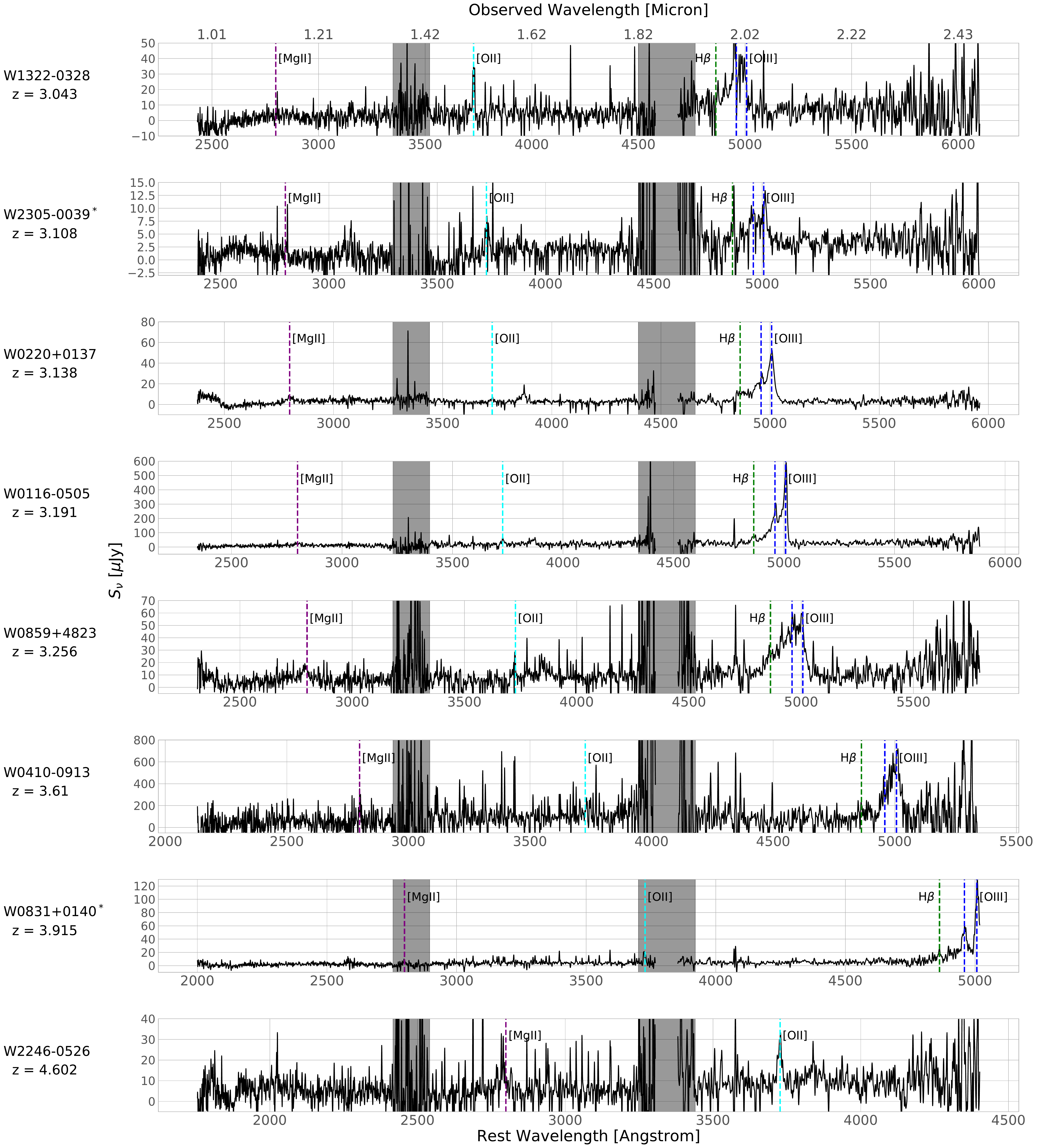}
    \label{halfdogs3}
\end{figure*}

\section{Spectral Analysis}\label{sec:extract}
\begin{figure*}
    \centering
    \noindent\includegraphics[width=38pc]{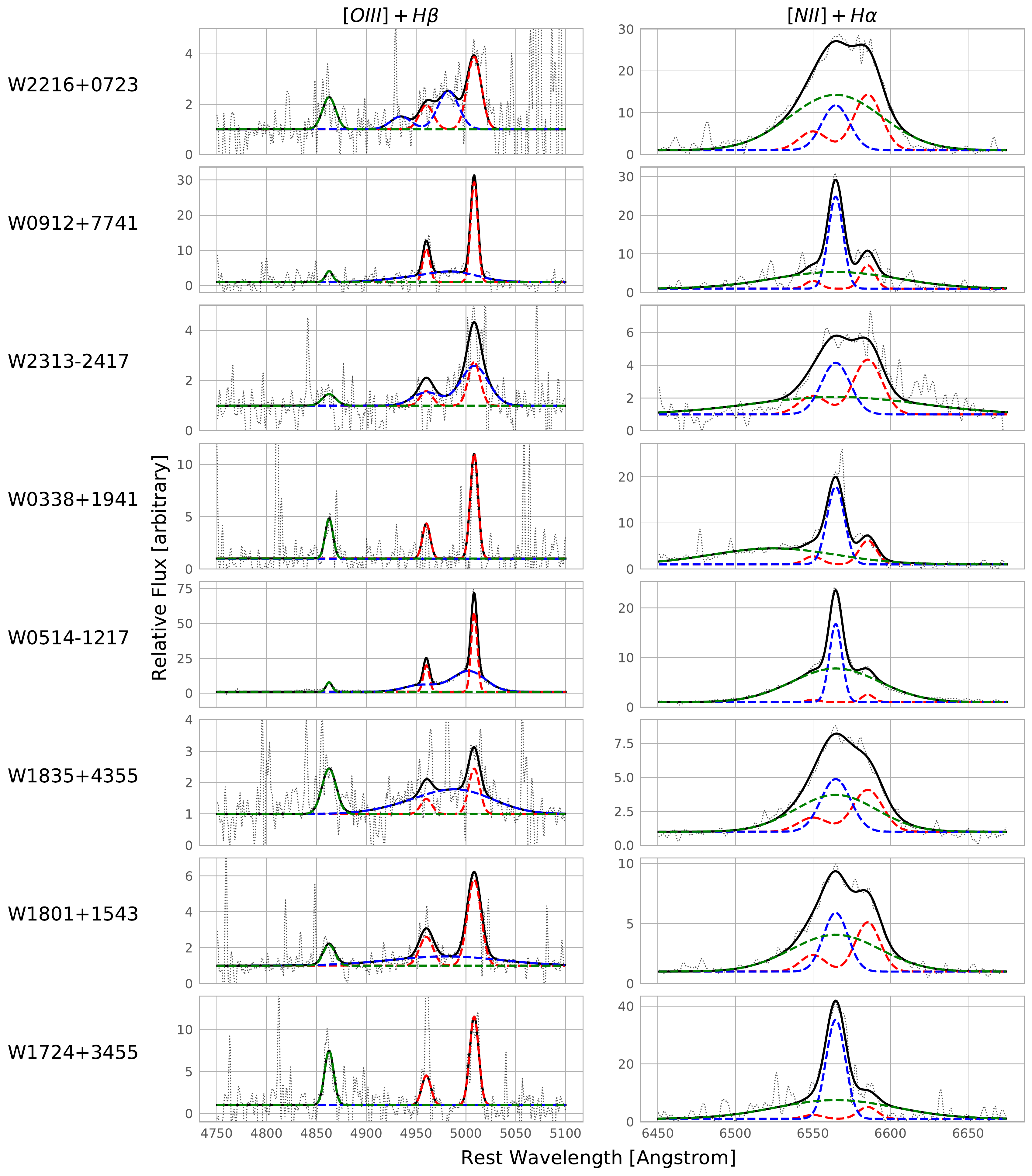}
    \caption{\oiii\ and \ha\ line fits by target, continuum-normalized. Targets are sorted by systemic redshift. In the \oiii\ plots, green represents \hb, blue represents broad/blueshifted \oiii, and red represents narrow/systemic \oiii. In the \ha\ plots, green represents broad \ha\, blue represents narrow \ha\, and red represents the \nii\ doublet. Black represents the total fit. Observed flux has been convolved with a two pixel Gaussian kernel for clarity and is plotted in dashed gray.  Fluxes are normalized to the local continuum by dividing a polynomial fit to the region around the line. Figure continues on the next page.}
    \label{lineplots1}
\end{figure*}
\begin{figure*}
    \centering
    \noindent\includegraphics[width=38pc]{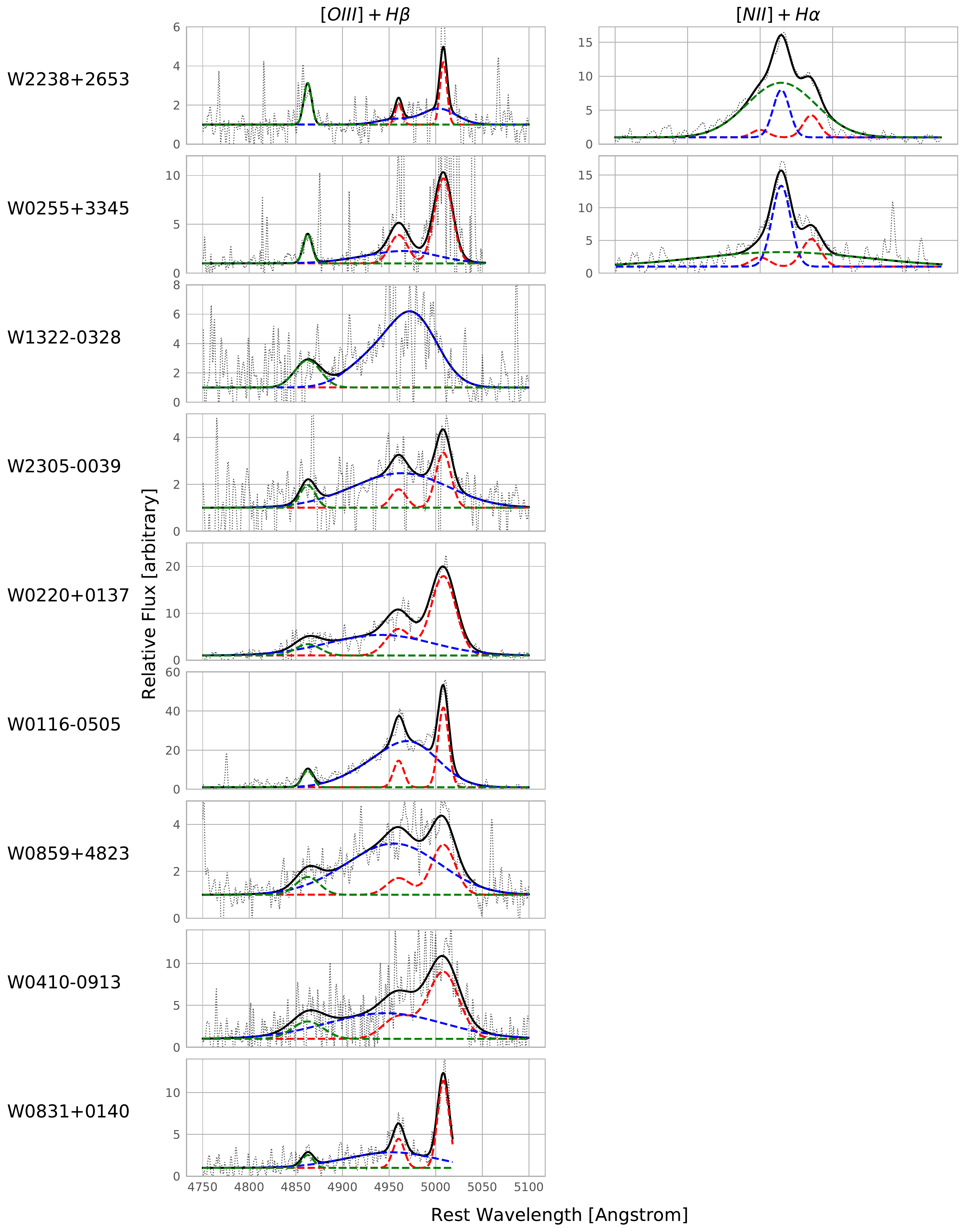}
    \label{lineplots2}
\end{figure*}
\subsection{Redshift Determinations}
Redshifts were determined by cross-correlating the full spectra with a template based on an SDSS emission line list \citep{SDSS} and are listed in Table \ref{obsprops}. The detection of multiple strong lines enables robust constraints for all targets except W0010+3236, W1905+5802, W1838+3429, and W2016-0041, with typical errors of $\sim10^{-3}$ in $z$. The redshift of 2.235 listed for W0514-1217 in Table \ref{obsprops} differs significantly from the spectroscopic value of 2.5 reported in \citet{Farrah_2017}. Other redshifts are in good agreement with previously-published values.  In addition to the redshift from cross-correlation, redshifts are also fit for individual lines, which typically agree with the cross-correlation value with a standard deviation $\sigma_z \approx 10^{-3}$ consistent with the resolution of NIRES and the error in the cross-correlation result. We therefore adopt the cross-correlation value as the systemic redshift for calculating velocities, and note that the \ha\ and narrow \oiii\ lines do not show significant velocity offsets from the systemic value.

W0010+3236, W1905+5802, and W2235+1605 lack previously published redshifts. The NIRES spectra of W0010+3236 and W1905+5802 show no significant features, preventing a determination. The weak features in the 2.3--2.4$\mu$m range are not statistically significant, and appear to be related to increased thermal noise at long wavelengths. These objects may be relatively nearby (z $<0.55$), based on the strong continuum detection compared with other sources, and the strong optical emission lines are therefore not observable with NIRES. W2235+1605 similarly lacks a published redshift, but a single strong line is detected near 1.43$\mu$m.  We believe this line is \oiii$\lambda5007$\r A at a redshift $z = 1.857$ based on the presence of a weak feature at the location which would correspond to \oiii$\lambda4959$\r A with the approximate expected intensity ratio, though strong telluric features prevent a clear detection. At $ z = 1.857$, \ha\ would fall in the gap between the $H$ and $K$-bands, and no other well-detected lines can be used to verify the redshift. We therefore do not include W2235+1605 in subsequent analysis. While W1838+3429 has a previously published redshift, the NIRES spectrum does not clearly detect any emission lines, despite the predicted presence of \oiii$\lambda 5007$ \r A at 2.1 $\mu$m based on the redshift from \citet{Tsai_2015}. Similarly, the NIRES spectrum cannot confirm the published redshift of 2.61 for W2016-0041 from \citet{Jun_2020}, which would correspond to \ha\ at 2.37 $\mu$m and \oiii\ at 1.81 $\mu$m. Finally, while the NIRES spectrum is consistent with the \citet{Jun_2020} redshift for W1719+0446, no individual line is sufficiently well-detected to obtain a good fit to the line profile, and we drop W1719+0446 from subsequent analysis.

Although the broad \oiii\ feature in W1322-0328 prefers a redshift $z = 3.025$, the [\ion{O}{2}]$\lambda3727$\r A feature indicates $z = 3.043$, which is consistent with the published in \citet{Tsai_2015}. We therefore adopt adopt $z = 3.043$ for W1322-0328, and constrain the \oiii\ fitting to this value.

The NIRES spectrum for W2246-0526 is consistent with the strong detection of [OII]$\lambda 3727$\r A and weak detection of [\ion{Mg}{2}]$\lambda2799$\r A at a redshift of 4.602, in good agreement with the values reported by \citet{Tsai_2018} from [\ion{C}{4}] and [\ion{Mg}{2}] detections and by \citet{diazsantos_2018} from $\rm Ly\alpha$. 

\subsection{Line Profile Fitting Routine}
All line profile parameters were fit using a custom-built Markov-Chain Monte Carlo (MCMC) fitting routine.  Prior to fitting each line, the flux was normalized by dividing a 4th-degree polynomial fit to the surrounding continuum. Each line or set of lines was fit with one or more Gaussian profiles in wavelength space. A single walker was initialized near the result of a $\chi^2$ minimization of the model line profile and run for $10^5$ steps, trimming the first five thousand to eliminate any remaining burn-in. An additional parameter was added to all models to account for the error in the observed spectrum, which was not well-determined from the reduction pipeline. This additional parameter does not affect the values of the best-fit line profile parameters, but is necessary to obtain accurate estimates for the fit errors. The median values of each parameter's MCMC chain are considered the best-fit line profile parameters, and 1$\sigma$ errors are determined by the upper and lower bounds which enclose 34 percent of the chain from the median.

\subsection{\texorpdfstring{\ha}{Ha} and \texorpdfstring{\nii}{[NII]} Profiles}
Figure \ref{lineplots1} shows significant blending between the \ha\ and \nii\ in all targets with detected \ha\ emission. We therefore constrain the \nii\ and \ha\ features to a single redshift. The \ha\ emission is fit with two Gaussians to fit both narrow and broad emission. The \nii$\lambda 6548,\lambda 6584$\r A doublet ratio is fixed to 0.338, and the \nii\ width is fixed to the narrow \ha\ width. This assumes the \nii\ and narrow \ha\ emission arise from similar physical environments with similar kinematics. No targets show evidence of emission from the nearby [\ion{Fe}{2}] line complex, which is therefore not included in the fitting. Figure \ref{lineplots1} plots the \ha\ profile fits in the right column, with \oiii\ plotted at left. The final \ha\ template consists of six free parameters for redshift of the line complex, \ha\ broad amplitude, \ha\ broad width, \nii\ amplitude, \ha\ narrow amplitude, and narrow \ha/\nii\ width. Line profile parameters are listed in Table \ref{velprops2}. The use of a single redshift for both broad and narrow \ha\ emission, with no outflow component, is discussed in more detail in Section \ref{sec:profskin}.

\begin{figure}
    \centering
    \noindent\includegraphics[width=20pc]{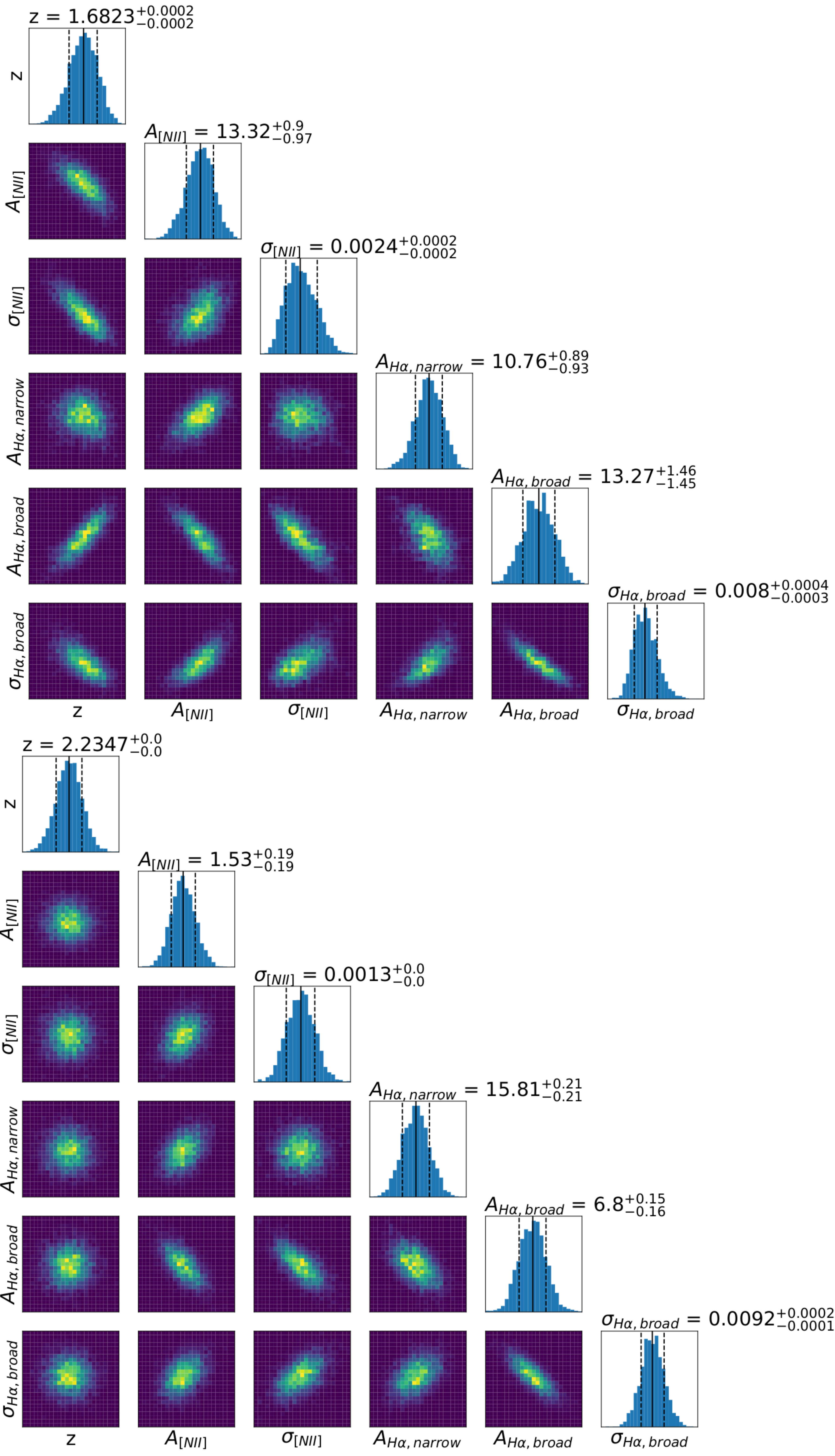}
    \caption{Corner plot from the \ha\ fit to W2216+0723 in the top panel and W0514-1217 at bottom.  Degeneracies between parameters are clearly present in W2216+0723, particularly for the redshift and the broad \ha\ components. This may allow significantly different kinematics to still offer a reasonable goodness-of-fit, beyond what is expected from our reported errors.  In contrast, the clearly resolved \nii\ emission in W0514-1217 breaks the degeneracies between \nii\ and broad \ha\, resulting in minimal covariance between fitting parameters and more reliable error estimation.}
    \label{corners}
\end{figure}

The blending of the \nii\ and \ha\ lines complicates the fitting and error estimation for some targets. The effect is most notable for W1835+4355, W2216+0723, and W2313-2417 and is present to a lesser extent in W1801+1543 and W2238+2653.  The blending of the line profile results in strong covariances between the redshift and line parameters, as shown in the top of Figure \ref{corners} for the case of W2216+0723.  These covariances can result in degenerate fitting.  Estimates of both the \nii/\ha\ ratio and its error are also less reliable in the presence of covariances. In cases where \nii\ is resolved separately, the degeneracies between parameters are broken, and more reliable fit results are obtained. This can be seen in the case of W0514-1217, plotted at the bottom of Figure \ref{corners}.

W0338+1941 displays a unique \ha\ line profile among the observed targets. The clear asymmetry in the line profile was best fit by allowing the line center of the broad \ha\ emission component to vary freely. While the kinematics differ substantially from the observed \oiii\ profile, which shows no broad/blueshifted emission, it is consistent with the published spectrum of \citet{Wu_2018} for W0338+1941. We discuss the origin of the broad \ha\ emission in Section \ref{sec:profskin}.

\subsection{\texorpdfstring{\oiii}{[OIII]} and \texorpdfstring{\hb}{Hb} Profiles}
Due to blending in some targets, we fit the \oiii\ and \hb\ lines simultaneously. For targets with detected \ha\ emission, we fix the \hb\ width to the narrow \ha\ width, as no {sources show a clear broad \hb\ component, presumably due to signal-to-noise limitations. For targets without detected \ha\ emission, the \hb\ width is fixed to the width of the narrow \oiii\ emission. The \oiii$\lambda4959,\lambda5007$\r A intensity ratio is fixed to 0.335. Fitting a single narrow component to the \oiii\ doublet resulted in a poor fit to most of the spectra, and therefore a second, blueshifted \oiii\ doublet was included with a free redshift. The resulting template consisted of seven free parameters for \oiii/\hb\ redshift, \hb\ amplitude, \oiii\ amplitude, \oiii\ width, blueshifted \oiii\ redshift, blueshifted \oiii\ amplitude, and blueshifted \oiii\ width. Line profiles parameters are listed in Table \ref{velprops1}.}

There is no statistically compelling evidence for a blueshifted \oiii\ component in W0338+1941 or W1724+3455, so a fit with only a single \oiii\ doublet and \hb\ was used. Similarly, W1322-0328 is best fit without a narrow/systemic component due to the width of the broad component and poor signal-to-noise, and is therefore fit with \hb\ and an outflowing \oiii\ doublet. The poor quality of the spectrum of W1322-0328 required significant additional constraints on the fitting to avoid fitting continuum features or telluric residuals, and the resulting \hb\ measurement should be considered an upper limit. All targets except W2313-2417 prefer a significant blueshift to the broad \oiii\ emission. As was the case in the \ha\ fitting, no targets show evidence of the [\ion{Fe}{2}] line complex, which was therefore not included. 

No clear \oiii\ detection is made in seven targets. A tentative detection is made in W2235+1605, but suffers from significant telluric contamination which prevents a clear identification. The redshift of W2246-0526 is too high to detect \oiii$\lambda$5007 \r A. W0010+3236 W1905+5802, and W2016-0041 have featureless NIRES spectra. W1719+0446 appears to have a small feature near the expected location of the \oiii\ doublet, but the quality of the spectrum is insufficient to obtain a fit. In W1838+3429, no \oiii\ detection is made despite a redshift from \citet{Tsai_2015} which would place the \oiii\ feature in the $K$-band.

\subsection{Other Lines}
In addition to \hb, \oiii$\lambda\lambda5007$\r A, \ha, and \nii$\lambda\lambda6548$\r A, detections of [\ion{O}{2}]$\lambda3727$\r A (13 targets), [\ion{O}{1}]$\lambda6302$\r A (4 targets), [\ion{O}{1}]$\lambda6363$\r A (W1801+1543), [\ion{Mg}{2}]$\lambda2799$\r A (2 targets), H$\delta$ (W0514-1217), [\ion{He}{1}]$\lambda3889$\r A (W0514-1217), and [\ion{Ne}{6}]$\lambda3427$\r A (W0514-1217) are listed in the appendix, along with \sii\ doublet detections (9 targets). For all but \sii, a Gaussian template with three free parameters was used to fit redshift, line width $\sigma$, and line intensity to the spectrum. For the \sii\ doublet, the redshift was assumed to match \ha\ and a three parameter template consisting of two Gaussians with a fixed wavelength offset was attempted to fit line width, intensity of the 6732\r A line, and the \sii$\lambda6732/\lambda6718$ ratio. Significant blending of the doublet components results in poor constraints on the line ratio in all targets except W0514-1217, limiting our ability to use the \sii\ feature to estimate electron densities, and most targets were better fit with a single Gaussian. We therefore report the total equivalent width for the doublet in Table \ref{otherlines}. 

\subsection{Extinction Estimates}\label{sub:extinct}
For the seven targets with both \ha\ and \hb\ detections, we estimate the optical extinction in Table \ref{ratioprops} by comparison with the intrinsic Balmer decrement.  The $A_{V,narrow}$ estimates compare the \hb\ flux with the narrow component of the \ha\ emission, while the $A_V,tot$ estimates use the total \ha\ emission including the broad component. The validity of these estimates is dependent on the origin of the broad component to the \ha\ emission. If a significant portion of the broad Balmer emission originates from a high-density AGN Broad-Line Region (BLR), the use of the total emission in estimating extinction will result in inaccuracies due to differences in the intrinsic Balmer decrement between the BLR and lower-density emitting regions \citep{osterbrock}. In this case, $A_{V,narrow}$ will provide a better estimate of the average extinction.
In contrast, an outflow origin for the broad \ha\ emission would have the same intrinsic Balmer decrement as the narrow component, resulting in $A_{V,tot}$ providing a good upper limit for the average extinction in the narrow line gas, as the \hb\ feature is fit with a single component due to the signal-to-noise of the spectra.

All targets except W1835+4355 and W2238+2653 have narrow Balmer decrements substantially above the extinction-free value of 3.1 \citep{Kim_2006}, indicative of significant extinction at optical wavelengths. Large optical extinctions in Hot DOGs were also reported in \citet{Assef_2015} and \citet{Jun_2020}. Applying the Balmer decrement/extinction relation from \citet{Dom_nguez_2013} with an intrinsic decrement of 3.1 to the measured \ha/\hb\ flux ratios gives a mean $A_{V,narrow}$ of 2.4 magnitudes and median of 3 magnitudes, disregarding W1835+4355. These estimates are likely to be lower limits, as the dust obscuration will result in a bias in the observed emission towards less-obscured regions. As expected, the $A_{V,tot}$ estimates are significantly larger, with a mean of 6.4 magnitudes and median of 6 magnitudes. While these estimates suffer from the same bias towards less-obscured regions, the ambiguity in the origin of the broad component will cause $A_{V,tot}$ to overestimate the true extinction if the broad component arises from an outflow which is not detected in the \hb\ line profile.  The unphysical $A_{V,narrow}$ value obtained for W1835+4355 and large errors on other targets are a result of the poor quality of the \hb\ detections in many targets.

Due to the possibility of differential extinction between emission components and the uncertainty in the origin of the observed broad \ha\ emission, correcting line luminosities based on either of the $A_V$ estimates in Table \ref{ratioprops} is unreliable. The large measured Balmer decrements suggest significant extinction is present at optical wavelengths, but assessing the precise impact of extinction of particular emission components would require spatially-resolved spectroscopy to clarify the origin of the broad \ha\ emission and address the possibility of differential extinction between the narrow/systemic and broad/blueshifted components to the \oiii\ emission.

\section{Results}\label{sec:results}
The results of the line extractions for the \hb/\oiii\ complex and \ha/\nii\ complex are presented in Tables \ref{ratioprops}, \ref{velprops1}, and \ref{velprops2}. Table \ref{ratioprops} presents the ratios of line fluxes and optical extinction estimates, while Table \ref{velprops1} presents the measured line profile parameters for \oiii\ and \hb. Table \ref{velprops2} presents the line profile parameters for \nii\ and \ha.  Figure \ref{bpt} plots the line ratios from Table \ref{ratioprops} on a Baldwin-Phillips-Terlevich (BPT) diagram \citep{bpt}. Fits to additional lines detected but not otherwise discussed are listed and plotted in the appendix.

\begin{deluxetable*}{ccccccccccc}\centering
\tablewidth{0pt}
\tabletypesize{\scriptsize}
\tablecaption{Line Ratios}
\tablehead{ & \colhead{$\frac{[OIII]\lambda 5007_{nar}}{H\beta}$} & \colhead{$\frac{[OIII]\lambda 5007_{br}}{H\beta}$} & $\frac{[OIII]\lambda 5007_{tot}}{H\beta}$ & $\frac{[OIII]\lambda 5007_{br}}{[OIII]\lambda 5007_{nar}}$& $\frac{[NII]\lambda 6585}{H\alpha_{tot}}$ & $\frac{H\alpha_{br}}{H\alpha_{nar}}$ & $\frac{H\alpha_{nar}}{H\beta}$ & $A_{V,nar}$ & $\frac{H\alpha_{tot}}{H\beta}$ & $A_{V,tot}$}
\startdata
W0116-0505 & 4.6$\pm$0.5 & 13$\pm$1 & 17$\pm$2 & 2.8$\pm$0.2 &  - & - &  - & - & - & - \\
W0220+0137 & 7$\pm$1 & 6$\pm$1.0 & 13$\pm$2 & 0.8$\pm$0.1 &  - & - &  - & - & - & - \\
W0255+3345 & 6$\pm$1 & 3$\pm$1 & 9$\pm$2 & 0.5$\pm$0.2 & 0.13$\pm$0.03 & 1.6$\pm$0.4 & 8$\pm$3 & 3$\pm$2 & 20$\pm$9 & 6$\pm$2 \\
W0338+1941 & 2.6$\pm$0.6 & - & 2.6$\pm$0.6 & - & 0.13$\pm$0.02 & 1.5$\pm$0.2 & 6$\pm$4 & 2$\pm$2 & 15$\pm$10 & 6$\pm$2 \\
W0410-0913 & 4$\pm$2 & 4$\pm$2 & 8$\pm$3 & 1.1$\pm$0.3 &  - & - &  - & - & - & - \\
W0514-1217 & 13.6$\pm$0.8 & 20$\pm$1 & 33$\pm$2 & 1.44$\pm$0.05 & 0.024$\pm$0.003 & 3.0$\pm$0.1 & 4.4$\pm$0.5 & 1.2$\pm$0.9 & 17$\pm$2 & 6.0$\pm$0.9 \\
W0831+0140 & 7$\pm$1 & 7$\pm$2 & 14$\pm$3 & 1.1$\pm$0.2 &  - & - &  - & - & - & - \\
W0859+4823 & 3$\pm$1 & 8$\pm$3 & 11$\pm$3 & 3.0$\pm$0.6 &  - & - &  - & - & - & - \\
W0912+7741 & 9$\pm$2 & 6$\pm$2 & 15$\pm$4 & 0.7$\pm$0.1 & 0.10$\pm$0.02 & 1.5$\pm$0.2 & 8$\pm$7 & 3$\pm$3 & 21$\pm$17 & 7$\pm$3 \\
W1322-0328$^*$ & - & 5$\pm$1 & 5$\pm$1 & - &  - & - &  - & - & - & - \\
W1724+3455 & 1.6$\pm$0.2 & - & 1.6$\pm$0.2 & - & 0.06$\pm$0.02 & 1.2$\pm$0.3 & 7$\pm$5 & 3$\pm$2 & 16$\pm$11 & 6$\pm$2 \\
W1801+1543 & 4.7$\pm$0.9 & 3$\pm$1 & 8$\pm$2 & 0.7$\pm$0.2 & 0.26$\pm$0.03 & 2.3$\pm$0.3 & 3.5$\pm$0.8 & 0.4$\pm$1 & 11$\pm$3 & 5$\pm$1 \\
W1835+4355 & 0.8$\pm$0.2 & 2.3$\pm$0.6 & 3.1$\pm$0.7 & 3$\pm$1 & 0.27$\pm$0.04 & 1.9$\pm$0.3 & 2.6$\pm$0.6 & -0.6$\pm$1.3 & 8$\pm$2 & 3$\pm$1 \\
W2216+0723 & 2.4$\pm$0.6 & 1.9$\pm$0.5 & 4$\pm$1& 0.8$\pm$0.2 & 0.24$\pm$0.04 & 4.1$\pm$0.7 & 10$\pm$5 & 4$\pm$2 & 50$\pm$20 & 10$\pm$2 \\
W2238+2653 & 1.3$\pm$0.4 & 1.7$\pm$0.7 & 3.0$\pm$0.9 & 1.2$\pm$0.6 & 0.08$\pm$0.01 & 4.8$\pm$0.6 & 3$\pm$1 & 0.3$\pm$1.7 & 20$\pm$8 & 6$\pm$2 \\
W2305-0039 & 2.4$\pm$0.7 & 7$\pm$2 & 10$\pm$3 & 2.9$\pm$0.6 &  - & - &  - & - & - & - \\
W2313-2417 & 3$\pm$3 & 7$\pm$5 & 10$\pm$7 & 2$\pm$2 & 0.36$\pm$0.07 & 2.0$\pm$0.5 & 15$\pm$8 & 5$\pm$2 & 40$\pm$30 & 9$\pm$2 \\
\enddata
\tablecomments{$^*$ indicates the \hb\ detection used in estimating the line ratios is an upper limit. The \oiii/\hb\ and $A_V$ values for these targets are thus lower limits. Optical extinction $A_V$ calculated from \citet{Dom_nguez_2013}, assuming an intrinsic Balmer decrement of 3.1. Poor detection of \hb\ leads to large errors in some line ratios.}
\label{ratioprops}
\end{deluxetable*}

\begin{deluxetable*}{cccccccccc}\centering
\tablewidth{0pt}
\tabletypesize{\scriptsize}
\tablecaption{\oiii\ and \hb\ Line Profile Parameters}
\tablehead{ & \colhead{$FWHM_{[OIII]}^{narrow}$} & \colhead{$EW_{[OIII]}^{narrow}$} & \colhead{$FWHM_{[OIII]}^{outflow}$} & \colhead{$EW_{[OIII]}^{outflow}$} & \colhead{$\Delta v_{broad-narrow}$} & \colhead{$FWHM_{H\beta}$} & \colhead{$EW_{H\beta}$} \\  & \colhead{(km s$^{-1}$)} & \colhead{(\r A)} & \colhead{(km s$^{-1}$)} & \colhead{(\r A)} &  \colhead{(km s$^{-1}$)} & \colhead{(km s$^{-1}$)} & \colhead{(\r A)} }
\startdata
W0116-0505 & 800$\pm$30 & 2400$\pm$100 & 4200$\pm$100 & 6700$\pm$300 & -2030$\pm$50 & 820$\pm$30 & 520$\pm$50 \\
W0220+0137 & 1880$\pm$40 & 2340$\pm$70 & 7300$\pm$400 & 1900$\pm$100 & -3400$\pm$200 & 1940$\pm$40 & 330$\pm$60 \\
W0255+3345 & 1400$\pm$200 & 800$\pm$100 & 5000$\pm$2000 & 300$\pm$200 & -200$\pm$1000 & 690$\pm$50 & 130$\pm$20 \\
W0338+1941 & 550$\pm$60 & 310$\pm$40 &  -  &  -  &  -  & 570$\pm$60 & 120$\pm$20 \\
W0410-0913$^*$ & 2300$\pm$100 & 1500$\pm$200 & 8300$\pm$900 & 1700$\pm$300 & -3000$\pm$1000 & 2400$\pm$100 & 400$\pm$100 \\
W0514-1217 & 430$\pm$10 & 1390$\pm$20 & 2350$\pm$50 & 2010$\pm$70 & -330$\pm$20 & 440$\pm$10 & 100$\pm$10 \\
W0831+0140 & 870$\pm$50 & 790$\pm$60 & 6000$\pm$1000 & 800$\pm$200 & -2800$\pm$400 & 890$\pm$50 & 120$\pm$20 \\
W0859+4823$^*$ & 1800$\pm$300 & 290$\pm$50 & 6400$\pm$500 & 870$\pm$90 & -2600$\pm$300 & 1800$\pm$300 & 100$\pm$30 \\
W0912+7741 & 510$\pm$20 & 760$\pm$40 & 3500$\pm$400 & 530$\pm$90 & -1300$\pm$200 & 520$\pm$20 & 80$\pm$20 \\
W1322-0328 &  -  &  -  & 3700$\pm$300 & 1300$\pm$200 & -2000$\pm$100 & 2000$\pm$300 & 260$\pm$60 \\
W1724+3455 & 640$\pm$40 & 400$\pm$40 &  -  &  -  &  -  & 670$\pm$30 & 250$\pm$30 \\
W1801+1543 & 970$\pm$70 & 270$\pm$20 & 7000$\pm$1000 & 180$\pm$50 & -1100$\pm$800 & 850$\pm$30 & 60$\pm$10 \\
W1835+4355 & 800$\pm$200 & 70$\pm$20 & 4900$\pm$800 & 200$\pm$50 & -800$\pm$400 & 1030$\pm$40 & 80$\pm$10 \\
W2216+0723 & 1000$\pm$200 & 140$\pm$30 & 1500$\pm$300 & 100$\pm$20 & -1500$\pm$100 & 960$\pm$80 & 56$\pm$9 \\
W2238+2653$^*$ & 530$\pm$120 & 100$\pm$30 & 2600$\pm$700 & 130$\pm$50 & -3000$\pm$400 & 610$\pm$40 & 75$\pm$9 \\
W2305-0039 & 1200$\pm$100 & 200$\pm$30 & 6600$\pm$800 & 590$\pm$90 & -2100$\pm$300 & 1200$\pm$100 & 80$\pm$20 \\
W2313-2417$^*$ & 900$\pm$500 & 80$\pm$60 & 2000$\pm$600 & 200$\pm$100 & 20$\pm$200 & 990$\pm$50 & 20$\pm$10 \\
\enddata
\tablecomments{Equivalent widths and FWHM for \oiii\ and \hb\ detections. W0338+1941 and W1724+3455 were best fit with a single \oiii\ component in the rest frame, while W1322-0328 was best fit with only an outflow \oiii\ component. Errors are based on the 1$\sigma$ confidence interval from the MCMC histogram, which may underestimate errors in the presence of significant covariances. $^*$ indicates multiple parameters were strongly degenerate in the MCMC corner plots. $\Delta v_{broad-narrow}$ is the velocity difference between the narrow/systemic \oiii\ emission and the broad/blueshifted \oiii\ component.}
\label{velprops1}
\end{deluxetable*}

\begin{deluxetable*}{cccccccccc}\centering
\tablewidth{0pt}
\tabletypesize{\scriptsize}
\tablecaption{\nii\ and \ha\ Line Profile Parameters}
\tablehead{ & \colhead{$FWHM_{[NII]}$} & \colhead{$EW_{[NII]}$} & \colhead{$FWHM_{H\alpha}^{narrow}$} & \colhead{$EW_{H\alpha}^{narrow}$} & \colhead{$FWHM_{H\alpha}^{broad}$} & \colhead{$EW_{H\alpha}^{broad}$} \\ & \colhead{(km s$^{-1}$)} & \colhead{(\r A)} & \colhead{(km s$^{-1}$)} & \colhead{(\r A)} & \colhead{(km s$^{-1}$)} & \colhead{(\r A)}}
\startdata
W0255+3345 & 690$\pm$50 & 250$\pm$40 & 690$\pm$50 & 730$\pm$60 & 6000$\pm$1000 & 1200$\pm$300 \\
W0338+1941$^{**}$ & 610$\pm$10 & 230$\pm$30 & 610$\pm$30 & 750$\pm$40 & 4300$\pm$400 & 1100$\pm$100 \\
W0514-1217 & 440$\pm$10 & 51$\pm$6 & 440$\pm$10 &   520$\pm$10 & 3050$\pm$50 & 1560$\pm$40 \\
W0912+7741 & 520$\pm$20 & 220$\pm$30 & 520$\pm$20 & 860$\pm$50 & 4300$\pm$400 & 1300$\pm$200 \\
W1724+3455 & 670$\pm$30 & 220$\pm$80 & 670$\pm$30 & 1800$\pm$100 & 4200$\pm$600 & 2100$\pm$500 \\
W1801+1543$^*$ & 850$\pm$30 & 270$\pm$20 & 860$\pm$30 & 330$\pm$20 & 3100$\pm$100 & 730$\pm$80 \\
W1835+4355$^*$ & 1030$\pm$40 & 240$\pm$30 & 1030$\pm$40 & 310$\pm$30 & 2800$\pm$100 & 590$\pm$90 \\
W2216+0723$^*$ & 960$\pm$80 & 800$\pm$90 & 960$\pm$80 & 650$\pm$80 & 3200$\pm$100 & 2600$\pm$300 \\
W2238+2653$^*$ & 610$\pm$40 & 160$\pm$20 & 610$\pm$40 & 340$\pm$30 & 2550$\pm$80 & 1600$\pm$100 \\
W2313-2417 & 990$\pm$50 & 230$\pm$30 & 990$\pm$50 & 220$\pm$20 & 5800$\pm$900 & 400$\pm$100 \\
\enddata
\tablecomments{Equivalent widths and FWHM for \nii\ and \ha\ detections. Propagated errors do not account for covariance between fit parameters, which is substantial in targets with more blended line profiles, particularly W1835+4355 and W2216+0723.$^*$ indicates multiple parameters were strongly degenerate in the MCMC corner plots. $^{**}$ the broad \ha\ emission in W0338+1941 is blueshifted by $1800\pm200$ \kms\ with respect to the narrow component, while in all other cases the broad \ha\ is at the same redshift as narrow \ha\ and \nii.}
\label{velprops2}
\end{deluxetable*}

\subsection{Star Formation vs AGN Activity}
\begin{figure*}
    \centering
    \noindent\includegraphics[width=39pc]{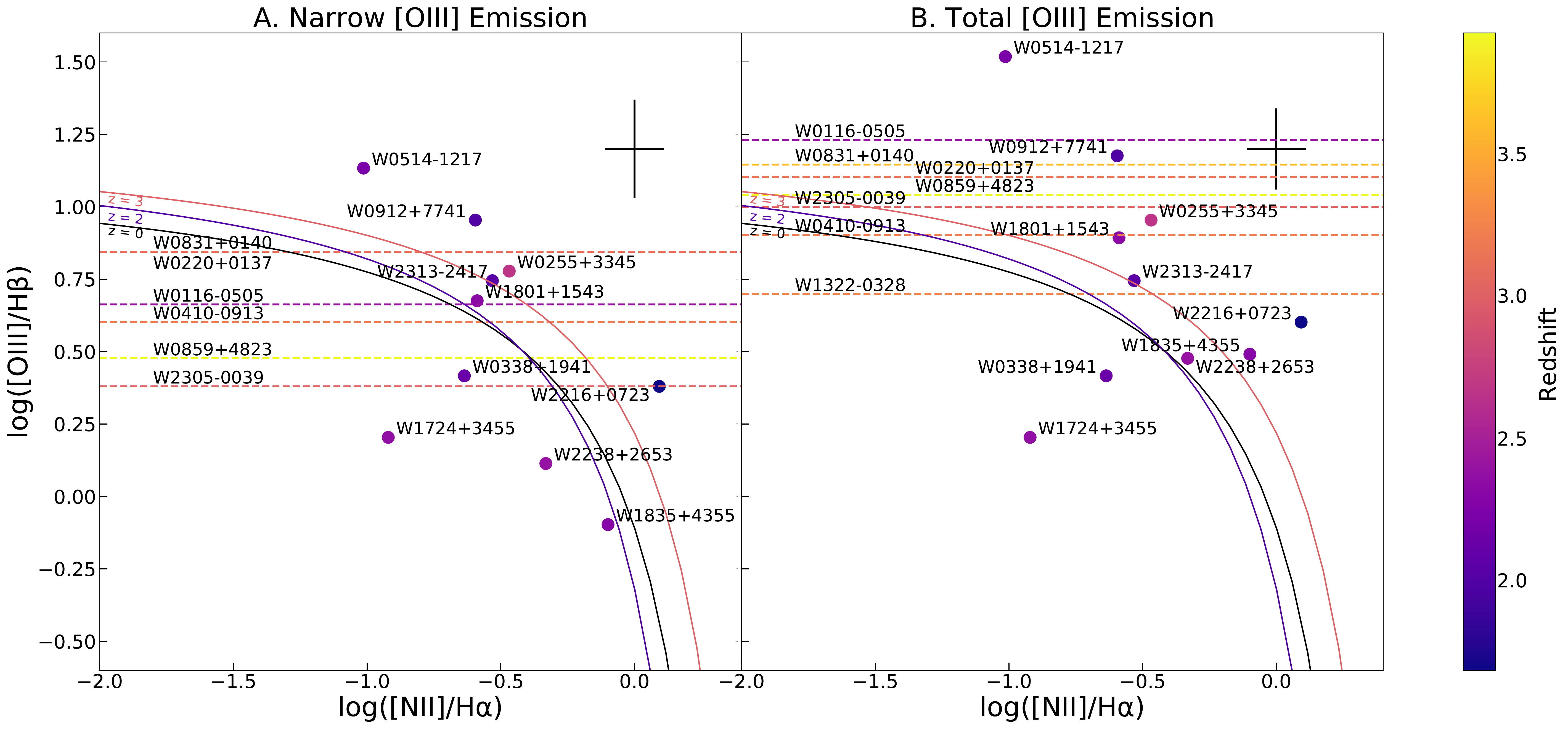}
    \caption{BPT diagram for all targets with measured \oiii.  At left, the narrow \oiii\ is used for the \oiii/\hb\ ratio. At right, the total \oiii\ is used for the \oiii/\hb\ ratio. In both cases only the narrow \ha\ is used for the  \nii/\ha\ flux ratio due to uncertainty in the origin of the broad component. Targets above log \oiii/\hb\ = 0.5 are likely to be AGN. The cross in the upper right of each panel indicates the typical measurement errors.  \ha\ is out of the wavelength range accessible to NIRES for $z > 2.7$, so targets with $z > 2.7$ are plotted as horizontal lines for the \oiii/\hb\ measurement. Star formation/AGN dividing contours are plotted for $z = 0$ \citep[][in black]{kewley_2001} as well as $z = 2$ and $z = 3$ \cite[][shaded by redshift]{Kewley_2013}, and points are shaded by the target redshift.}
    \label{bpt}
\end{figure*}
All targets with measured \oiii\ emission listed in Table \ref{ratioprops} are plotted on a BPT diagram \citep{bpt}, allowing us to explore the relative importance of star formation and AGN activity in the observed Hot DOGs.  Diagrams were made using both the total \oiii\ flux (Figure \ref{bpt}, right panel) and only the the flux from the narrow/systemic component (Figure \ref{bpt}, left panel). The \nii/\ha\ ratios plotted in Figure \ref{bpt} uses the narrow component of the \ha\ fit. The width of the \hb\ emission is fixed to match the value obtained from narrow \ha\ where available, or match the narrow \oiii\ when no \ha\ detection is made. The use of the narrow \ha\ emission and fitting \hb\ with the same width should prevent the ambiguity in the origin of the broad component from significantly impacting placements on the BPT diagram.

\citet{Tsai_2015} found the bolometric luminosity of Hot DOGs is AGN-dominated. However, the $\rm 10^{14}\ L_\odot$ luminosities of Hot DOGs are large enough that even a significant star formation component would be difficult to distinguish based on the infrared SED. In the case of the Hot DOG W1814+3412, \citet{Eisenhardt_2012} estimates an extinction-corrected star formation rate of 300 $\rm M_\odot\ yr^{-1}$, despite star formation accounting for $<10$ percent of the bolometric luminosity. 

Of the 17 targets presented here with detected \oiii\ emission, at least four, and possibly as many as eight appear to have a significant starburst component based on the BPT diagram placement, in addition to broad \ha\ emission indicative of AGN activity. W0338+1941 and W1724+3455 lack broad \oiii\ emission and are unambiguously in the star-formation dominated region, while W2238+2653 and W1835+4355 are near the starburst/AGN transition, with the classification depending on the inclusion of the broad/blueshifted \oiii\ emission in the \oiii/\hb\ ratio. W2305-0039, and W0859+4823 may also be star-formation dominated based on the narrow/systemic \oiii/\hb\ ratio, though the high redshifts placing \ha\ beyond the reach of NIRES prevents a precise placement on the BPT. The remaining targets are placed in the AGN-dominated region, though W0410-0913 and W0116-0505 are ambiguous due to the lack of an \ha\ detection. 

The increase in \oiii/\hb\ from the inclusion of the broad/blueshifted \oiii\ component is consistent with either an AGN-driven or a shock-driven origin for the broad/blueshifted \oiii\ emission. The two scenarios could be distinguished if a broad component to the \nii\ emission analogous to the \oiii\ broad/blueshifted component could be identified. However, the blending of the \nii\ and \ha\ profiles makes such a distinction impossible in the available Hot DOG spectra, and an acceptable fit is obtained in all targets with only a single \nii\ component. Furthermore, the extreme ($>10$) \oiii/\hb\ ratios in some Hot DOGs are inconsistent with shock models \citep{Rich_2010}, and the high ionization suggests the presence of an AGN.

While the selection criteria and presence of large ionized outflows suggest all targets host luminous AGN, several sources appear to be starburst dominated in the narrow-component BPT diagram (Left panel, Figure \ref{bpt}). However, outflow properties do not seem to correlate with location on the BPT diagram, and both the \oiii\ and \ha\ line profiles are broadly similar across all targets. The large $A_V$ values in Table \ref{ratioprops} and the similarity in the observed spectra suggests the differences in location on the BPT diagram may be a result of dust obscuration. In highly-obscured objects the rest-frame optical spectra may not readily detect a buried AGN, and may instead by dominated by star formation in the host galaxy. In such cases, an AGN may still be detectable at longer wavelengths, and previous rest-frame infrared observations of Hot DOGs are consistent with dust heating by AGN \citep{Wu_2012}. Due to the high obscuration towards the AGN component, the fraction of AGN-dominated sources among the Hot DOGs as revealed by their rest-frame optical spectra should therefore be considered a lower limit on the true AGN-dominated fraction. The underestimate of the \oiii/\hb\ ratios due to the inclusion of some broad \hb\ emission by the single-component fit further biases the BPT placements towards the star-formation dominated region.

\subsection{Line Profiles and Gas Kinematics}\label{sec:profskin}
Nearly all (15/17) targets with detected \oiii\ emission show evidence of significant broad and blueshifted \oiii\ emission in addition to narrower emission at the systemic redshift determined from other lines.  Ratios of line fluxes are presented in Table \ref{ratioprops} and kinematic properties of lines are reported in Tables \ref{velprops1} and \ref{velprops2}. Typical blueshift velocities of the centroid of the broad/blueshifted component compared with the narrow/systemic component are listed in Table \ref{velprops1} and are on the order of 2000 km/s. The FWHM of this broad component tends to be substantially larger, on the order of 4000--5000 km/s.  The relative contribution of the blueshifted component to the overall \oiii\ emission varies dramatically. W0859+4823 provides a clear example of the blueshifted emission dominating the total \oiii\ luminosity, contributing more than 80 percent of the total line flux. However, in W0220+0137, W0255+3345, W0912+7741, and W2216+0723, the majority of the \oiii\ luminosity is in the narrow component, and W0338+1941 and W1724+3455 entirely lack a broad/blueshifted component, suggesting significant variation in the properties of the broad component between objects. Broadened asymmetric blueshifted line profiles are indicative of ionized outflows in nearby Seyferts \citep[e.g.][]{osterbrock, Schmidt_2018}, though the blueshifted components in those objects do not dominate the total emission as is the case in the Hot DOGs. 

The \nii\ and \ha\ lines do not show similar kinematics to the broad/blueshifted \oiii\ emission. Attempts to match the \ha\ emission profile with the kinematics of the broad/blueshifted \oiii\ emission resulted in a statistically worse fit ($\rm \Delta BIC > 10$). For all targets except W0338+1941, the broad \ha\ component is consistent with the redshift of the narrow \ha\ component and \nii. This suggests that a significant fraction of the broad \ha\ emission may come from a high-density broad line region, and we therefore exclude this component from the \nii/\ha\ ratio used in Figure \ref{bpt}.

However, the strong blending of \ha\ and \nii\ lines in many targets means we cannot rule out a lower-velocity outflow in those lines, similar to the results from other Hot DOGs in \citet{Wu_2018}. AGN-driven \ha\ outflows with lower velocities than seen in \oiii\ are common in nearby Type 2 quasars \citep{kang_2017}, and may be present in W2216+0723, W1724+3455, and W2238+2653 (see Figure \ref{lineplots1}, right column). High velocity \oiii\ emission with no corresponding \ha\ has also been seen in other luminous, red galaxies at $z\approx2-3$ \citep{zakamska_2016}. Despite the possibility of weak outflows in \ha, fixing the broad \ha\ to the redshift of the narrow component resulted in good fits for all sources except W0338+1941. The use of a single component to fit the broad \ha\ emission despite the possible presence of outflows means we cannot be certain of the physical origin of the broad emission, or of the relative contributions of the broad-line region and outflows to the total \ha\ emission.

W0338+1941 is the only Hot DOG in the sample to show clear evidence of a fast outflow in the \ha\ profile. While the \oiii\ profile does not show any broad/blueshifted component, the width and blueshift of the broad \ha\ component are 4300 \kms\ and 1800 \kms\, respectively, comparable to the broad \oiii\ kinematics seen in other Hot DOGs. The lack of an observed outflow in \oiii\ suggests the broad/blueshifted emission is due to \ha\ rather than the \nii\ doublet, which has a comparable critical density to the \oiii\ doublet. This suggests W0338+1941 hosts an outflow similar to those seen in other Hot DOGs, but at a higher density and/or lower ionization which results in the outflow being visible in the \ha\ profile and not \oiii, in contrast with other Hot DOGs. Future spatially-resolved spectroscopy may offer insight into how the outflow in W0338+1941 differs from other Hot DOGs.

\section{Discussion}\label{sec:disc}
\subsection{Outflow Energies}\label{sec:outnrg}
\begin{deluxetable*}{ccccccc}\centering
\tablewidth{0pt}
\tabletypesize{\scriptsize}
\tablecaption{Outflow Luminosities, Energies, and Masses}
\tablehead{ & \colhead{$\rm L_{outflow}$} & \colhead{$\rm M_{outflow}$} &  \colhead{$\rm \dot{M}_{outflow}$} & \colhead{$\rm v_{outflow}$} & \colhead{$\rm \dot{E}_{out}$} & \colhead{$\rm P_{out}$} \\ & $\rm \log L_\odot$ &  $\rm \log M_\odot$ & $\rm \log M_\odot\ yr^{-1}$ & \kms & $\rm \log erg\ s^{-1}$ & $\rm \log dyn$ }
\startdata
W0116-0505 & $11.5\pm0.1$ & $9.2\pm0.3$ & $3.9\pm0.3$ & $5400\pm100$ &  $46.9\pm0.3$ & $38.5\pm0.3$ \\
W0220+0137 & $9.8\pm0.1$ & $7.5\pm0.3$ & $2.5\pm0.3$ & $9300\pm300$ &  $45.9\pm0.3$ & $37.3\pm0.3$ \\
W0255+3345 & $9.2\pm0.3$ & $6.9\pm0.4$ & $1.7\pm0.4$ & $6000\pm1000$ &  $44.7\pm0.4$ & $36.3\pm0.4$ \\
W0410-0913 & $10.6\pm0.2$ & $8.3\pm0.4$ & $3.3\pm0.4$ & $9000\pm1000$ &  $46.8\pm0.4$ & $38.1\pm0.4$ \\
W0514-1217 & $10.4\pm0.1$ & $8.1\pm0.3$ & $2.5\pm0.3$ & $2100\pm100$ &  $44.6\pm0.3$ & $36.6\pm0.3$ \\
W0859+4823 & $10.0\pm0.1$ & $7.7\pm0.4$ & $2.6\pm0.3$ & $7600\pm400$ &  $45.9\pm0.3$ & $37.3\pm0.3$ \\
W0912+7741 & $9.4\pm0.3$ & $7.1\pm0.4$ & $1.8\pm0.4$ & $4000\pm400$ &  $44.5\pm0.4$ & $36.2\pm0.4$ \\
W1322-0328 & $10.0\pm0.2$ & $7.7\pm0.4$ & $2.4\pm0.3$ & $5000\pm200$ & $45.3\pm0.4$ & $36.9\pm0.4$ \\
W1801+1543 & $8.8\pm0.1$ & $6.5\pm0.3$ & $1.3\pm0.3$ & $6000\pm1000$ & $44.4\pm0.4$ & $35.9\pm0.3$ \\
W1835+4355 & $9.1\pm0.1$ & $6.8\pm0.3$ & $1.5\pm0.3$ & $4400\pm700$ & $44.3\pm0.4$ & $35.9\pm0.4$ \\
W2216+0723 & $8.8\pm0.2$ & $6.5\pm0.4$ & $1.0\pm0.4$ & $3300\pm200$ &  $43.6\pm0.4$ & $35.4\pm0.4$ \\
W2313-2417 & $8.4\pm0.3$ & $6.1\pm0.4$ & $0.3\pm0.4$ & $1700\pm600$ & $42.2\pm0.5$ & $34.3\pm0.5$ \\
\enddata
\tablecomments{Outflow properties estimated following \citet{Jun_2020}, using the relations from \citet{nesvadba_2011} and \citet{carniani_2015} with $R_{out} = 3$ kpc and $n_e = 300$ cm$^{-3}$ to derive the gas mass. Outflow luminosity is the luminosity of the broad/blueshifted component to the \oiii\ emission. Outflow velocity is estimated using \citet{Bae_2016, Bae_2017} to correct for projection effects. A filled spherical geometry is assumed \citep{maiolino_2012}.  These values should be considered strict lower limits, as as no reddening correction has been applied to the \oiii\ luminosity. W0831+0140, W2238+2653, and W2305-0039 are omitted due to issues with flux calibration, despite the presence of significant outflows. Error estimates for the mass outflow rate, energy injection rate, and outflow pressure have had an additional factor of two uncertainty added in quadrature in order to reflect the uncertainties associated with the outflow size and density assumptions.}
\label{outprops}
\end{deluxetable*}

Interpreting the broad/blueshifted \oiii\ emission as an ionized outflow, we estimate the energetic properties of the outflow. The \sii\ doublet ratio is well-constrained only for W0514-1217, with a best-fit value of $1.05\pm0.15$. In the remaining targets, the lines of the \sii\ doublet are too broad to obtain a reliable estimate for the intensity ratio, and the doublet is fit with a single Gaussian. The ratio from W0514-1217 corresponds to an electron density $\rm n_e \approx 300\ cm^{-3}$ \citep{draine_2003}. While we adopt this value for subsequent calculations, the variation in the centroids for targets with blended \sii\ doublets suggests a range of densities among Hot DOGs. Spatially-resolved observations of outflows find electron densities in nearby AGN outflows range from $10^2-10^3 \rm\ cm^{-3}$ \citep[e.g.][]{greene_2011, harrison_2014, Karouzos_2016}, broadly consistent with our assumed value of $\rm 300\ cm^{-3}$. To calculate the mass of outflowing gas, we adopt the relations from \citet{nesvadba_2011} and \citet{carniani_2015}:
\begin{equation}
M_{gas} = 4\times10^7\ M_\odot \times \left( \frac{L_{[OIII],outflow}}{10^{44} \textrm{ erg s}^{-1}} \right) \left( \frac{\langle n_e \rangle}{10^3 \textrm{cm}^{-3}} \right)^{-1}
\end{equation}
Where $L_{[OIII],outflow}$ is the luminosity of the broad/blueshifted component to the \oiii\ emission and we have assumed $\langle n_e \rangle^2 / \langle n_e^2 \rangle \times 10^{-[O/H]} \approx 1$ such that the \citet{nesvadba_2011} and \citet{carniani_2015} relations are equivalent \citep[see][for details]{Jun_2020}. We next calculate the effective outflow velocity, correcting for projection and dust extinction effects for a spherical geometry \citep{Bae_2016, Bae_2017}:
\begin{equation}
   v_{out} = 2\sqrt{\sigma_{[OIII], broad}^2 + \Delta v_{[OIII], broad}^2}
\end{equation}
Where $\sigma_{[OIII], broad}$ is the standard deviation of the Gaussian fit to the broad/blueshifted \oiii\ component and $\Delta v_{[OIII], broad}$ is the shift of the broad/blueshifted \oiii\ component relative to the narrow/systemic component. From the mass and velocity of the outflow, we can then define mass ejection rate, energy injection rate, and momentum flux as follows, assuming a filled spherical geometry \citep[e.g.][]{maiolino_2012}:
\begin{equation}
\begin{split}
\dot{M}_{out} & = \frac{3 M_{gas} v_{out}}{R_{out}}\\
    \dot{E}_{out} & = \frac{3 M_{gas} v_{out}^3}{2 R_{out}}\\
    P_{out} & = \frac{3 M_{gas} v_{out}^2}{R_{out}}\\
\end{split}
\end{equation}

The resulting estimates for outflow properties are listed in Table \ref{outprops}. As the NIRES spectra are not spatially resolved, the outflow size must be assumed. We use $R_{out} \approx 3 \rm kpc$, as in \citet{Jun_2020}. This choice is motivated by spatially-resolved observations of nearby outflows, which are typically $\sim 1-10 \rm\ kpc$ \citep[e.g.][]{harrison_2014, Karouzos_2016a, Kang_2018}. As no reddening/extinction corrections have been applied and only the ionized component of the outflow is considered, the values in Table \ref{outprops} may be lower limits to the true outflow properties. Outflow mass, mass outflow rate, kinetic energy, kinetic energy injection rate, and momentum flux are linearly proportional to the broad/blue \oiii\ luminosity. If the extinctions estimated from the Balmer ratio are applicable to the outflow see in \oiii, the derived outflow masses may be underestimated by a factor of $\sim2$. While assumptions were required for both the size and density of the outflow, the derived outflow properties depend linearly on these assumptions, and the density and size are likely to be inversely correlated, partially mitigating these assumptions on the final derived quantities. To reflect the additional of these assumptions, the error estimates in Table \ref{outprops} include an additional factor of two uncertainty added in quadrature to the uncertainty in the luminosity arising from the flux calibration.

We measure mass outflow rates up to $\rm \sim8000\ M_\odot\ yr^{-1}$ (median 150 $\rm M_\odot\ yr^{-1}$, mean 950 $\rm M_\odot\ yr^{-1}$), and typical energy injection rates on the order of $\rm 10^{45}\ erg\ s^{-1}$.  These estimates are similar to those reported in \citet{Jun_2020}, which range from $\sim60$ to $\sim4300\ \rm M_\odot\ yr^{-1}$, median 970 $\rm M_\odot\ yr^{-1}$, with typical energy injection rates on the order of $\rm 10^{45}\ erg\ s^{-1}$. W2216+0723 was also observed in \citet{Jun_2020}, allowing a direct comparison of the analyses. Despite similar outflow velocities, we report a dramatically smaller mass outflow rate. The difference appears to be a result of differences in the flux calibration. We use the broad-band photometry from \citet{Assef_2015} to perform flux calibration, which underestimates the flux compared with the spectrum presented in \citet{Jun_2020}. We caution that flux calibration in faint targets can be highly uncertain, and further note the presence of strong telluric features near the \oiii\ line in W2216+0723 which may interfere in flux calibration for this target specifically. 

The mass outflow rate for W0116-0505 is larger than that of any other observed Hot DOG, while the velocities of the outflows in W0410-0913 and W0220+0137 are the largest observed. There does not appear to be a significant connection between the outflow velocities and outflow luminosities, though the possibility of varying levels of extinction between objects may hide any such correlation. The largest outflows are found in higher-redshift targets, though the present sample size is too small to make a robust connection. We discuss the comparison of the Hot DOG outflows to local sources in Section \ref{sec:feedback}.

\subsection{Star Formation Rate Estimates}\label{sub:sfrs}
\begin{deluxetable}{ccccc}\centering
\tablewidth{0pt}
\tabletypesize{\scriptsize}
\tablecaption{Star Formation Rates}
\tablehead{ & \colhead{$\rm L_{H\alpha,narrow}$} & \colhead{$\rm SFR_{Balmer}$}  & \colhead{$\rm SFR_{Balmer,corr}$} & \colhead{$\rm SFR_{IR}$} \\ & $\rm \log L_\odot$ & $\rm M_\odot\ yr^{-1}$  & $\rm M_\odot\ yr^{-1}$ & $\rm M_\odot\ yr^{-1}$} 
\startdata
W0116-0505$^*$ & 10.8 & 1300 & - & 2700  \\
W0338+1941 & 8.8  & 10 &  80 & 2600 \\
W0410-0913$^*$ & 10.5 & 70 & - & 5300  \\
W1801+1543 & 9.0 & 20 & 30 & 2600 \\
W0859+4823$^*$ & 9.6 & 80 & - & 2600 \\
W1724+3455 & 10.0 & 200 & 3200 & 1100  \\
W1835+4355 & 9.1 & 30 & 30 & 1800 \\
W2238+2653 & - & -  & - & 2700 \\
W2305-0039$^*$ & 9.1 & 30  & - & 2000 \\
\enddata
\tablecomments{Star formation rate estimates for sources which appear star-formation dominated in Figure \ref{bpt}. $\rm SFR_{Balmer}$ uses the uncorrected narrow Balmer line luminosity and assumes case B recombination. Objects marked $^*$ estimate the \ha\ luminosity from \hb\ assuming case B and no extinction. $\rm SFR_{IR}$ use the relation from \citet{Murphy_2011}, estimating the star-formation proportion from the \citet{Kirkpatrick_2015} SED library with an additional hot dust component (see text for details). $\rm SFR_{Balmer,corr}$ corrects the observed \ha\ luminosities using $A_{V,nar}$ estimates from Table \ref{ratioprops}. The NIRES spectrum for W2238+2653 is not well flux-calibrated (see Section \ref{sec:obs}), so there is no optical star formation rate estimate for this source.}
\label{sfrtab}
\end{deluxetable}
Figure \ref{bpt} indicates at least four, and possibly as many as eight, of the Hot DOGs have optical emission line ratios consistent with heating by young stars. Due to the extreme bolometric luminosities of Hot DOGs, these sources could have very high star formation rates, even if star formation accounts for only a small proportion of the total luminosity. For targets with a Balmer line detection and where the narrow component indicates significant star formation from the BPT diagram, we can estimate the star formation rate assuming case B recombination
 \citep{osterbrock_2006, Murphy_2011}:
\begin{equation}
    \rm \left(\frac{SFR_{Balmer}}{M_\odot\ yr^{-1}}\right) = 5.37\times10^{-42}\left(\frac{L_{H\alpha}}{erg\ s^{-1}}\right)
\end{equation}
Where $\rm L_{H\alpha}$ is the luminosity of the narrow \ha\ component, which we used to identify the star-forming Hot DOGs in the BPT diagram. For targets without observations of \ha, we estimate $\rm L_{H\alpha}$ from $\rm L_{H\beta}$ assuming case B recombination. We use both the uncorrected value of $\rm L_{H\alpha}$ and the value after applying the $A_{V,nar}$ estimates from Table \ref{ratioprops}. We exclude W0255+3345, W0514-1217, W0912+07741, W1801+1543, W2216+0713, and W2313-2417 from the $\rm SFR_{Balmer}$ calculation based on the BPT placement in the left panel of Figure \ref{bpt}. W2238+2653 did not have an accurate absolute flux calibration and the $\rm SFR_{Balmer}$ could not be calculated, despite falling in the star-forming region of the BPT. The uncorrected Balmer-line SFRs have a mean of 250 and median of 50 $\rm M_\odot\ yr^{-1}$. The uncertainties are expected to be on the order of 20--30 percent, dominated by the continuum flux rather than the line equivalent width. The errors on the extinction correction add an additional factor of $\sim2$ uncertainty to the corrected Balmer SFR for W1835+4355 and a factor of $\sim6$ for W1724+3455. Values are listed in Table \ref{sfrtab}.

Since Hot DOGs are inherently dusty galaxies and the star formation rates derived from the rest-frame optical emission lines may be lower limits, even when corrected for extinction, we also calculate a SFR from the cold dust emission for comparison. For each Hot DOG in Table \ref{sfrtab}, we fit the available MIR/FIR photometry with the composite SED library from \citet{Kirkpatrick_2015} to derive the fraction of the total $\rm 8-1000\ \mu m$ luminosity contributed by star formation. Since the templates do not contain galaxies with the hot dust excess characteristic of Hot DOGs, we add a black body dust emission component with a temperature similar to the 450 K emission identified in \citet{Tsai_2015}. We assume that this hot dust emission is entirely powered by a central AGN. For all objects the AGN contribution to the MIR flux is at least 90 percent, and is typically 60--70 percent for the total 8--1000 $\mu$m emission. After deriving the total fractional IR emission contributed by star formation, we use the relation in \citet{Murphy_2011} to estimate the corresponding star formation rates:
\begin{equation}
    \rm \left(\frac{SFR_{IR}}{M_\odot\ yr^{-1}}\right) = 3.88\times10^{-44}\left(\frac{L_{IR}}{erg\ s^{-1}}\right)
\end{equation}
The resulting values for each object are listed in  the last column of Table \ref{sfrtab} rounded to the nearest 100 $\rm M_\odot\ yr^{-1}$, and have a mean and median of 2600 $\rm M_\odot\ yr^{-1}$. The spacing of the model grid suggests these values are accurate to within about ten percent. However, because the SED fitting procedure consistently prefers the template with the one of the largest $\rm 8-1000\ \mu m$ $f_{\rm AGN}$ values in the model grid, the star formation rate values should be seen as upper limits. We note that while applying the narrow extinction correction to the $\rm SFR_{Balmer}$ estimate for W1724+3455 gives a corrected Balmer value much larger than $\rm SFR_{IR}$, the uncertainty in the optical extinction correction for this source is large.

The Balmer-derived SFRs in our Hot DOG sample are comparable to main-sequence galaxies with stellar masses $\rm M_* \approx 10^{10} - 10^{11} M_\odot$ at $z\sim2-3$ \citep{Daddi_2007,elbaz_2011, Speagle_2014}. Without a careful estimate of the stellar masses, best made by modelling the rest-frame near-infrared continuum, it is difficult to determine the location of our star-forming Hot DOGs relative to the main sequence at $z\sim2$. Low stellar masses would suggest these galaxies are experiencing a large burst of star formation, together with enhanced accretion onto the supermassive black hole. Large stellar masses, as have been reported for some Hot DOGs \citep[e.g.,][]{Assef_2015}, would place these galaxies on or below the main sequence, suggesting they may be in the act of quenching star formation.

\subsection{Black Hole Masses and Eddington Ratios}
\begin{deluxetable}{cccccc}\centering
\tablewidth{0pt}
\tabletypesize{\scriptsize}
\tablecaption{Black hole masses and Eddington ratios}
\tablehead{ & \colhead{$\sigma_{gas}$} & ${M_{BH,\sigma}}$ &  $L_{Edd}$ & $L_{bol}$ & $f_{Edd}$ \\ &  \kms & $\log M_\odot$ & $\log L_\odot$ & $\log L_\odot$ & }
\startdata
W0116-0505 & $340\pm10$ & $9.5$ & 14.0 & 14.1 & 1 \\
W0220+0137 & $600\pm200$ & 10.6 & 15.1 & 13.9 & 0.06 \\
W0255+3345 & $290\pm20$ & 9.2 & 13.7 & 13.0 & 0.2 \\
W0338+1941 & $240\pm30$ & $8.8$ & 13.4 & 13.3 & 1 \\
W0410-0913 & $990\pm50$ & $11.5$ & 16.0 & 14.2 & 0.02 \\
W0514-1217 & $181\pm2$ & $8.3$ & 12.8 & 14.1 & 20 \\
W0831+0140 & $370\pm20$ & $9.7$ & 14.2 & 14.3 & 1 \\
W0859+4823$^*$ & $260\pm90$ & 9.0 & 13.5 & 14.5 & 10 \\
W0912+7741 & $210\pm10$ & $8.6$ & 13.1 & 13.3 & 2 \\
W1322-0328 & $900\pm100$ & $11.3$ & 15.8 & 14.5 & 0.05 \\
W1724+3455$^*$ & $270\pm20$ & $9.1$ & 13.6 & 13.7 & 1 \\
W1801+1543 & $410\pm30$ & $9.9$ & 14.5 & 13.9 & 0.3  \\
W1835+4355$^*$ & $330\pm70$ & $9.4$ & 13.9 & 13.9 & 1 \\
W2216+0723 & $430\pm80$ & $9.9$ & 14.4 & 13.1 & 0.05 \\
W2238+2653$^*$ & $230\pm50$ & $8.7$ & 13.2 & 14.0 & 6 \\
W2305-0039$^*$ & $490\pm50$ & $10.2$ & 14.7 & 14.3 & 0.4 \\
W2313-2417 & $400\pm200$ & $9.6$ & 14.1 & 13.5 & 0.3 \\
\enddata
\tablecomments{$\sigma_{gas}$ will overestimate $\sigma_*$ and therefore all $M_{BH}$ values are upper limits. Values larger than $10^{10}\rm\ M_\odot$ are likely due to non-gravitational broadening of emission lines. $L_{bol}$ values are re-stated from Table \ref{obsprops}, and were calculated with a conservative power-law interpolation over MIR/FIR photometry as described in \citet{Tsai_2015}. Objects marked $^*$ have optical line ratios consistent with a star-forming galaxy, which would invalidate the $f_{Edd}$ estimation. }
\label{bhprops}
\end{deluxetable}
SED fitting to MIR/FIR photometry indicates all of the observed Hot DOGs are AGN-dominated. The apparently star-formation dominated objects in Figure \ref{bpt} are likely the result of significant dust extinction obscuring the central AGN, allowing the less-extincted star formation luminosity to dominate at rest-frame optical wavelengths, while the AGN continues to dominate in the infrared. We will therefore estimate black hole masses and Eddington ratios for all observed Hot DOGs, regardless of optical classification, under the assumption that Hot DOGs are AGN with varying levels of obscuration at optical wavelengths. Upper limits on the black hole mass are estimated $M_{BH} - \sigma_*$ relation from \citet{Kormendy_2013}, using the narrow-line emission width as an upper limit on $\sigma_*$:
\begin{multline}
\log{\frac{M_{BH}}{10^9 M\odot}} < -(0.510\pm0.049)+\\
    (4.377\pm0.290)\log{\frac{\sigma_*}{200 km/s}}
\end{multline}
Where $\sigma_*$ is the stellar velocity dispersion, which we replace with the $\sigma$ value determined from the narrowest detected emission feature. Narrow-line regions of AGN have been known to introduce significant line broadening beyond the Keplerian motion (e.g., AGN shocks or outflows), with emission lines overestimating $\sigma_*$ by up 50--100 percent, resulting in an overestimate of the black hole masses $M_{BH,\sigma}$ \citep{Bennert_2018}. We minimize the effect of this broadening by using the narrowest observed emission feature, which should be least impacted by non-gravitational motion and therefore provide a better estimate for $\sigma_*$. It is possible some of the larger $M_{BH}$ estimates (those well above $10^10\rm\ M_\odot$ as in W1322-0328) are overestimates due to non-gravitational gas motions, in particular in cases where $\sigma_{gas} > 400$ \kms, the largest $\sigma_*$ seen in inactive galaxies, but we report them here in an effort to calculate the masses consistently across the sample. Using the resulting upper limits on $M_{BH}$, we then calculate lower limits on the Eddington ratios, using the conservative power-law bolometric luminosity estimates from Table \ref{obsprops}:
\begin{equation}
\begin{split}
    L_{Edd} = 3.2 * 10^4 * \left(\frac{M_{BH}}{M_\odot}\right) L_\odot \\
    f_{Edd} = \frac{L_{bol}}{L_{Edd}} \\
\end{split}
\end{equation}

We note that upper limits for the black hole masses should result in lower limits on the Eddington ratio, as the power-law method used to estimate the bolometric luminosity from MIR/FIR photometry in Table \ref{obsprops} from \citet{Tsai_2015} provides a lower limit. Despite this, nine out of the seventeen Hot DOGs for which we make estimates have $f_{Edd} \geq 1$, and three more have $f_{Edd} \geq 0.3$, indicating super-Eddington accretion is common in Hot DOGs. Similar findings for Hot DOG Eddington ratios are presented in \citet{Wu_2018}, \citet{Tsai_2018}, and \citet{Jun_2020}. While the radiation pressure from super-Eddington accretion provides a possible mechanism for driving the observed outflows, the lack of observed outflows in W1724+3455 despite $f_{Edd} \sim 1$ and the presence of such outflows in W0255+3345 and W2216+0723 despite a low value of $f_{Edd}$ suggests \oiii\ outflows may not be uniquely linked to the accretion rate.

The calculation for $f_{Edd}$ assumed $L_{bol} \approx L_{AGN}$, which assumes the star-forming component does not contribute significantly to the overall luminosity. We believe this is a reasonable assumption, as the $f_{\rm AGN}$ estimates made earlier do not account for the emission from AGN-heated dust which dominates the Hot DOG SEDs in the MIR \citep{Tsai_2015}. Furthermore, as the $L_{bol}$ values underestimate the true luminosity, the presence of a $\sim10$ percent star-formation contribution to the total luminosity would not necessarily cause a significant difference between the values listed in Table \ref{bhprops} and the true $f_{Edd}$. We also note that the 0.3 dex intrinsic scatter of the $M_{BH}-\sigma_*$ relation introduces significant uncertainty in the Eddington ratio estimates due to the uncertainty in $M_{BH}$, though on average the $M_{BH,\sigma}$ listed in Table \ref{bhprops} should overestimate the true $M_{BH}$ due to the use of gas dispersion as an upper-limit estimate for $\sigma_*$. 

\subsection{Feedback on Host Galaxy}\label{sec:feedback}
The extreme mass outflow rates and outflow velocities in Table \ref{outprops} suggest that the central AGN is producing significant feedback on the host galaxy in Hot DOGs. The momentum fluxes listed in Table \ref{outprops} are generally of order $10L_{bol}/c$. Simulations of AGN-driven winds in galaxy mergers indicate the observed outflow velocities and momentum fluxes in Hot DOGs are sufficient to quench star formation over a few hundred Myr, and may also be capable of unbinding a substantial portion of the host galaxy's initial gas \citep{debuhr_2012}. Despite this, we observe star formation in Hot DOGs comparable to massive main-sequence galaxies at similar redshift, even without correcting \ha\ luminosities for extinction. The significant ongoing star formation suggests a more complicated interaction between the outflows and the ISM of the host galaxy, e.g., that the radial or angular extent of ionized outflows may not quench the entire host galaxy, or that the outflows are in the process of quenching star formation \citep[e.g.][]{hopkins_2012, woo_2017}. 

Extreme feedback is consistent with previous published observations of Hot DOGs. \citet{diazsantos_2016} obtained spatially-resolved observations of the [\ion{C}{2}] line in W2246-0526 which indicated a turbulent ISM and significant isotropic mass ejection.  Radiation pressure from super-Eddington accretion may expel significant amounts of material from the central AGN region, and possibly the host galaxy \citep{Assef_2015}, quenching star formation in the process. This would be consistent with the observation of extended Ly$\alpha$ emission near hot, dusty WISE sources at similar redshift \citep{bridge_2013}. The broad component kinematics in Table \ref{velprops1} and the energetics in Table \ref{outprops} are consistent with the explanation of Hot DOGs as a short-lived phase of intense accretion onto the central AGN coexisting with the expulsion of material due to strong feedback \citep[see][]{diazsantos_2018}.  

Massive outflows have also been observed in nearby ($z<1$) spatially-resolved ULIRGS. Observations of nearby ULIRGS have reported molecular outflows fom CO and OH observations with typical speeds of several hundred \kms\ and mass outflow rates of several hundred $\rm M_\odot\ yr^{-1}$, ranging up to $\sim1700\ \rm km\quad s^{-1}$ and $\sim1500\ \rm M_\odot\ yr^{-1}$ \citep{Gonz_lez_Alfonso_2017, Gowardhan_2018}. Observations of ionized lines in local ULIRGs have found similar outflow speeds but smaller mass outflow rates, on the order of 10 $\rm M_\odot\ yr^{-1}$ \citep{Soto_2012}, indicating that most of the mass is in the molecular component of the outflow.  At somewhat higher redshift (z = 1.4), ionized outflows with speeds up to 1700 \kms\ and mass outflow rates of 500--1500 $\rm M_\odot\ yr^{-1}$ have been reported for the quasar 3C298 \citep{Vayner_2017}. These mass outflow rates are similar to the values for Hot DOGs listed in Table \ref{outprops}. With limited spatial resolution, it is not possible to conclusively link the observed fast outflow in our sample Hot DOGs with the central AGN. However, all galaxies in this sample are AGN-dominated based on MIR photometry, which should be less impacted by dust obscuration compared with optical indicators, and the outflow speeds are well above any seen in starburst galaxies in the local Universe \citep{Heckman_1990}. the Hot DOG outflows are comparable to those seen in AGN-dominated ULIRGs, where the speed correlates with AGN power \citep{spoon_2009, Veilleux_2013}, and in some high-$z$ quasars as described above. We therefore believe the link between the observed fast outflows and AGN in the Hot DOGs is supported by the available observational evidence.

The dusty nature of Hot DOGS may imply that much of the outflow, at least in the central regions, remains hidden from view. Despite this, we find median outflow speeds and mass rates in excess of the most extreme ionized outflows in local ULIRGs, and comparable to high-redshift quasars. The mass outflow rates in the ionized gas are comparable to or greater than the derived uncorrected star formation rates, suggesting significant mass loading in the winds. When reddening and extinction effects are considered, ionized outflows in Hot DOGs may be substantially stronger than analogous features in other objects, and are likely to have significant feedback on the host galaxy.

\section{Summary and Conclusions}\label{sec:conc}
We obtained Keck2/NIRES spectra covering 0.95-2.42$\mu$m for a total of 21 Hot DOGs and three additional objects with a similar 4.6 to 22 $\mu$m SED. From the spectra and fits to the observed emission lines, we find:
\begin{enumerate}
    \item We obtain emission line redshifts for 20 objects ranging from $z = $1.7--4.6. No clear features were present in the spectra of four objects, preventing a redshift determination of the NIRES spectrum. Nine targets have $z > 3$, significantly expanding the number of high-redshift Hot DOGs with optical spectroscopy.
    
    \item  The \oiii$\lambda 5007$ line was detected in 17 objects.  Of these, 15 required the presence of a broad, blueshifted component. In 9 targets, the broad blueshifted component comprised the majority of the total [OIII] luminosity. These line profiles are indicative of massive, ionized outflows, with a median outflow rate of 150 $\rm M_\odot\ yr^{-1}$, and a maximum of 8000 $\rm M_\odot\ yr^{-1}$. These mass loss rates are significantly larger than those seen in ULIRGS in the local Universe, but comparable to those seen in some $z\sim2$ QSOs.
    
    \item Fits to the mid and far-infrared SEDs of Hot DOGs suggest they are all AGN-dominated. The presence of AGN is further supported by the detection of a broad component to the \ha\ emission in all ten targets where \ha\ is detected (FWHM of 2550-6200 \kms) at the systemic redshift. The broad \ha\ kinematics are distinct from the broad/blueshifted \oiii\ emission. Estimates of the Eddington ratios suggest accretion at or above the Eddington limit is common in Hot DOGs.
    
    \item Based on the rest-frame optical emission line flux ratios, we find evidence for vigorous, ongoing star formation in four and possibly as many as eight Hot DOGs, corresponding to 20--50 percent of the sample, despite the AGN dominating the total luminosity. The median star formation rates estimated from Balmer lines, uncorrected for reddening, is 50 $\rm M_\odot\ yr^{-1}$, with a range of 30--1300 $\rm M_\odot\ yr^{-1}$, comparable to that found in massive galaxies at $z\sim2-3$. The presence of powerful AGN, fast, massive outflows, and ongoing star formation may indicate that Hot DOGs are in a transition phase of rapid stellar mass and central black hole growth before feedback clears the nuclei of gas and dust and star formation is fully quenched. 
    
\end{enumerate}

\acknowledgments
We thank the anonymous reviewer for their helpful suggestions to improve this paper. The data presented herein were obtained at the W. M. Keck Observatory, which is operated as a scientific partnership among the California Institute of Technology, the University of California and the National Aeronautics and Space Administration. The Observatory was made possible by the generous financial support of the W. M. Keck Foundation. We wish to acknowledge the critical importance of the current and recent Mauna Kea Observatories daycrew, technicians, telescope operators, computer support, and office staff employees, especially during the challenging times presented by the COVID-19 pandemic.  Their expertise, ingenuity, and dedication is indispensable to the continued successful operation of these observatories.

This publication makes use of data products from the Wide-field Infrared Survey Explorer, which is a joint project of the University of California, Los Angeles, and the Jet Propulsion Laboratory/California Institute of Technology, funded by the National Aeronautics and Space Administration.

DSM was supported in part by a Leading Edge Fund from the Canadian Foundation 
for Innovation (project No. 30951) and a Discovery
Grant (RGPIN-2019-06524) from the Natural Sciences and Engineering Research Council (NSERC) of Canada. 
DSM is also grateful to the Dunlap Institute for Astronomy and Astrophysica at the University of Toronto for its contribution to the NIRES H2RG detector. T.D-S. acknowledges support from the CASSACA and CONICYT fund CAS-CONICYT Call 2018. C.-W. Tsai was supported by a grant from the NSFC (No. 11973051). This research was supported by the Basic Science Research Program through the National Research Foundation of Korea (NRF) funded by the Ministry of Education (NRF-2017R1A6A3A04005158).

The authors wish to recognize and acknowledge the very significant cultural role and reverence that the summit of Maunakea has always had within the indigenous Hawaiian community.  We are most fortunate to have the opportunity to conduct observations from this mountain.

\facilities{Keck:II (NIRES)}
\software{astropy (The Astropy Collaboration 2013, 2018)}
{\footnotesize
\bibliography{sample63}}

\begin{thebibliography}{}
\expandafter\ifx\csname natexlab\endcsname\relax\def\natexlab#1{#1}\fi
\providecommand{\url}[1]{\href{#1}{#1}}
\providecommand{\dodoi}[1]{doi:~\href{http://doi.org/#1}{\nolinkurl{#1}}}
\providecommand{\doeprint}[1]{\href{http://ascl.net/#1}{\nolinkurl{http://ascl.net/#1}}}
\providecommand{\doarXiv}[1]{\href{https://arxiv.org/abs/#1}{\nolinkurl{https://arxiv.org/abs/#1}}}

\bibitem[{Ahumada {et~al.}(2019)Ahumada, Prieto, Almeida, Anders, Anderson,
  Andrews, Anguiano, Arcodia, Armengaud, Aubert, Avila, Avila-Reese, Badenes,
  Balland, Barger, Barrera-Ballesteros, Basu, Bautista, Beaton, Beers,
  Benavides, Bender, Bernardi, Bershady, Beutler, Bidin, Bird, Bizyaev, Blanc,
  Blanton, Boquien, Borissova, Bovy, Brandt, Brinkmann, Brownstein, Bundy,
  Bureau, Burgasser, Burtin, Cano-Diaz, Capasso, Cappellari, Carrera,
  Chabanier, Chaplin, Chapman, Cherinka, Chiappini, Choi, Chojnowski, Chung,
  Clerc, Coffey, Comerford, Comparat, da~Costa, Cousinou, Covey, Crane, Cunha,
  da~Silva~Ilha, Dai, Damsted, Darling, Darrington, Jr., Davies, Dawson, De,
  de~la Macorra, Lee, de~Andrade~Queiroz, Machado, de~la Torre, Dell'Agli,
  du~Mas~des Bourboux, Diamond-Stanic, Dillon, Donor, Drory, Duckworth, Dwelly,
  Ebelke, Eftekharzadeh, Eigenbrot, Elsworth, Eracleous, Erfanianfar,
  Escoffier, Fan, Farr, Fernandez-Trincado, Feuillet, Finoguenov, Fofie,
  Fraser-McKelvie, Frinchaboy, Fromenteau, Fu, Galbany, Garcia,
  Garcia-Hernandez, Oehmichen, Ge, Maia, Geisler, Gelfand, Goddy, Goff,
  Gonzalez-Perez, Grabowski, Green, Grier, Guo, Guy, Harding, Hasselquist,
  Hawken, Hayes, Hearty, Hekker, Hogg, Holtzman, Hou, Hsieh, Huber, Hunt,
  Chitham, Imig, Jaber, Angel, Johnson, Jones, Jonsson, Jullo, Kim, Kinemuchi,
  IV, Kite, Klaene, Kneib, Kollmeier, Kong, Kounkel, Krishnarao, Lacerna, Lan,
  Lane, Law, Leung, Lewis, Li, Lian, Lin, Long, Longa-Pena, Lundgren, Lyke,
  Mackereth, MacLeod, Majewski, Manchado, Maraston, Martini, Masseron, Masters,
  Mathur, McDermid, Merloni, Merrifield, Meszaros, Miglio, Minniti, Minsley,
  Miyaji, Mohammad, Mosser, Mueller, Muna, Munoz-Gutierrez, Myers, Nadathur,
  Nair, do~Nascimento, Nevin, Newman, Nidever, Nitschelm, Noterdaeme,
  O'Connell, Olmstead, Oravetz, Oravetz, Osorio, Pace, Padilla,
  Palanque-Delabrouille, Palicio, Pan, Pan, Parker, Paviot, Peirani, Ramrez,
  Penny, Percival, Perez-Fournon, Perez-Rafols, Petitjean, Pieri, Pinsonneault,
  Poovelil, Povick, Prakash, Price-Whelan, Raddick, Raichoor, Ray, Rembold,
  Rezaie, Riffel, Riffel, Rix, Robin, Roman-Lopes, Roman-Zuniga, Rose, Ross,
  Rossi, Rowlands, Rubin, Salvato, Sanchez, Sanchez-Menguiano, Sanchez-Gallego,
  Sayres, Schaefer, Schiavon, Schimoia, Schlafly, Schlegel, Schneider,
  Schultheis, Schwope, Seo, Serenelli, Shafieloo, Shamsi, Shao, Shen, Shetrone,
  Shirley, Aguirre, Simon, Skrutskie, Slosar, Smethurst, Sobeck, Sodi, Souto,
  Stark, Stassun, Steinmetz, Stello, Stermer, Storchi-Bergmann, Streblyanska,
  Stringfellow, Stutz, Suarez, Sun, Taghizadeh-Popp, Talbot, Tayar, Thakar,
  Theriault, Thomas, Thomas, Tinker, Tojeiro, Toledo, Tremonti, Troup, Tuttle,
  Unda-Sanzana, Valentini, Vargas-Gonzalez, Vargas-Magana, Vazquez-Mata, Vivek,
  Wake, Wang, Weaver, Weijmans, Wild, Wilson, Wilson, Wolthuis, Wood-Vasey,
  Yan, Yang, Yeche, Zamora, Zarrouk, Zasowski, Zhang, Zhao, Zhao, Zheng, Zheng,
  Zhu, \& Zou}]{SDSS}
Ahumada, R., Prieto, C.~A., Almeida, A., {et~al.} 2019, The Sixteenth Data
  Release of the Sloan Digital Sky Surveys: First Release from the APOGEE-2
  Southern Survey and Full Release of eBOSS Spectra.
\newblock \doarXiv{1912.02905}

\bibitem[{Assef {et~al.}(2015)Assef, Eisenhardt, Stern, Tsai, Wu, Wylezalek,
  Blain, Bridge, Donoso, Gonzales, Griffith, \& Jarrett}]{Assef_2015}
Assef, R.~J., Eisenhardt, P. R.~M., Stern, D., {et~al.} 2015, The Astrophysical
  Journal, 804, 27, \dodoi{10.1088/0004-637x/804/1/27}

\bibitem[{Bae \& Woo(2016)}]{Bae_2016}
Bae, H.-J., \& Woo, J.-H. 2016, The Astrophysical Journal, 828, 97,
  \dodoi{10.3847/0004-637x/828/2/97}

\bibitem[{Bae {et~al.}(2017)Bae, Woo, Karouzos, Gallo, Flohic, Shen, \&
  Yoon}]{Bae_2017}
Bae, H.-J., Woo, J.-H., Karouzos, M., {et~al.} 2017, The Astrophysical Journal,
  837, 91, \dodoi{10.3847/1538-4357/aa5f5c}

\bibitem[{{Baldwin} {et~al.}(1981){Baldwin}, {Phillips}, \& {Terlevich}}]{bpt}
{Baldwin}, J.~A., {Phillips}, M.~M., \& {Terlevich}, R. 1981, \pasp, 93, 5,
  \dodoi{10.1086/130766}

\bibitem[{Bennert {et~al.}(2018)Bennert, Loveland, Donohue, Cosens, Lewis,
  Komossa, Treu, Malkan, Milgram, Flatland, Auger, Park, \&
  Lazarova}]{Bennert_2018}
Bennert, V., Loveland, D., Donohue, E., {et~al.} 2018, Monthly Notices of the
  Royal Astronomical Society, 481, 138, \dodoi{10.1093/mnras/sty2236}

\bibitem[{{Bridge} {et~al.}(2013){Bridge}, {Blain}, {Borys}, {Petty},
  {Benford}, {Eisenhardt}, {Farrah}, {Griffith}, {Jarrett}, {Lonsdale},
  {Stanford}, {Stern}, {Tsai}, {Wright}, \& {Wu}}]{bridge_2013}
{Bridge}, C.~R., {Blain}, A., {Borys}, C. J.~K., {et~al.} 2013, \apj, 769, 91,
  \dodoi{10.1088/0004-637X/769/2/91}

\bibitem[{{Carniani} {et~al.}(2015){Carniani}, {Marconi}, {Maiolino},
  {Balmaverde}, {Brusa}, {Cano-D{\'\i}az}, {Cicone}, {Comastri}, {Cresci},
  {Fiore}, {Feruglio}, {La Franca}, {Mainieri}, {Mannucci}, {Nagao}, {Netzer},
  {Piconcelli}, {Risaliti}, {Schneider}, \& {Shemmer}}]{carniani_2015}
{Carniani}, S., {Marconi}, A., {Maiolino}, R., {et~al.} 2015, \aap, 580, A102,
  \dodoi{10.1051/0004-6361/201526557}

\bibitem[{{Chary} \& {Elbaz}(2001)}]{chary_2001}
{Chary}, R., \& {Elbaz}, D. 2001, \apj, 556, 562, \dodoi{10.1086/321609}

\bibitem[{{Cushing} {et~al.}(2004){Cushing}, {Vacca}, \& {Rayner}}]{spex}
{Cushing}, M.~C., {Vacca}, W.~D., \& {Rayner}, J.~T. 2004, \pasp, 116, 362,
  \dodoi{10.1086/382907}

\bibitem[{{Cutri} \& {et al.}(2012)}]{Cutri_2012}
{Cutri}, R.~M., \& {et al.} 2012, VizieR Online Data Catalog, II/311

\bibitem[{Daddi {et~al.}(2007)Daddi, Dickinson, Morrison, Chary, Cimatti,
  Elbaz, Frayer, Renzini, Pope, Alexander, Bauer, Giavalisco, Huynh, Kurk, \&
  Mignoli}]{Daddi_2007}
Daddi, E., Dickinson, M., Morrison, G., {et~al.} 2007, The Astrophysical
  Journal, 670, 156, \dodoi{10.1086/521818}

\bibitem[{DeBuhr {et~al.}(2012)DeBuhr, Quataert, \& Ma}]{debuhr_2012}
DeBuhr, J., Quataert, E., \& Ma, C.-P. 2012, Monthly Notices of the Royal
  Astronomical Society, 420, 2221, \dodoi{10.1111/j.1365-2966.2011.20187.x}

\bibitem[{{D{\'\i}az-Santos} {et~al.}(2016){D{\'\i}az-Santos}, {Assef},
  {Blain}, {Tsai}, {Aravena}, {Eisenhardt}, {Wu}, {Stern}, \&
  {Bridge}}]{diazsantos_2016}
{D{\'\i}az-Santos}, T., {Assef}, R.~J., {Blain}, A.~W., {et~al.} 2016, \apjl,
  816, L6, \dodoi{10.3847/2041-8205/816/1/L6}

\bibitem[{{D{\'\i}az-Santos} {et~al.}(2018){D{\'\i}az-Santos}, {Assef},
  {Blain}, {Aravena}, {Stern}, {Tsai}, {Eisenhardt}, {Wu}, {Jun}, {Dibert},
  {Inami}, {Lansbury}, \& {Leclercq}}]{diazsantos_2018}
---. 2018, Science, 362, 1034, \dodoi{10.1126/science.aap7605}

\bibitem[{Dom{\'{\i}}nguez {et~al.}(2013)Dom{\'{\i}}nguez, Siana, Henry,
  Scarlata, Bedregal, Malkan, Atek, Ross, Colbert, Teplitz, Rafelski, McCarthy,
  Bunker, Hathi, Dressler, Martin, \& Masters}]{Dom_nguez_2013}
Dom{\'{\i}}nguez, A., Siana, B., Henry, A.~L., {et~al.} 2013, The Astrophysical
  Journal, 763, 145, \dodoi{10.1088/0004-637x/763/2/145}

\bibitem[{{Draine}(2011)}]{draine_2003}
{Draine}, B.~T. 2011, {Physics of the Interstellar and Intergalactic Medium}
  ({Princeton University Press})

\bibitem[{Eisenhardt {et~al.}(2012)Eisenhardt, Wu, Tsai, Assef, Benford, Blain,
  Bridge, Condon, Cushing, Cutri, Neal J.~Evans, Gelino, Griffith, Grillmair,
  Jarrett, Lonsdale, Masci, Mason, Petty, Sayers, Stanford, Stern, Wright, \&
  Yan}]{Eisenhardt_2012}
Eisenhardt, P. R.~M., Wu, J., Tsai, C.-W., {et~al.} 2012, The Astrophysical
  Journal, 755, 173, \dodoi{10.1088/0004-637x/755/2/173}

\bibitem[{{Elbaz} {et~al.}(2011){Elbaz}, {Dickinson}, {Hwang},
  {D{\'\i}az-Santos}, {Magdis}, {Magnelli}, {Le Borgne}, {Galliano},
  {Pannella}, {Chanial}, {Armus}, {Charmandaris}, {Daddi}, {Aussel}, {Popesso},
  {Kartaltepe}, {Altieri}, {Valtchanov}, {Coia}, {Dannerbauer}, {Dasyra},
  {Leiton}, {Mazzarella}, {Alexander}, {Buat}, {Burgarella}, {Chary}, {Gilli},
  {Ivison}, {Juneau}, {Le Floc'h}, {Lutz}, {Morrison}, {Mullaney}, {Murphy},
  {Pope}, {Scott}, {Brodwin}, {Calzetti}, {Cesarsky}, {Charlot}, {Dole},
  {Eisenhardt}, {Ferguson}, {F{\"o}rster Schreiber}, {Frayer}, {Giavalisco},
  {Huynh}, {Koekemoer}, {Papovich}, {Reddy}, {Surace}, {Teplitz}, {Yun}, \&
  {Wilson}}]{elbaz_2011}
{Elbaz}, D., {Dickinson}, M., {Hwang}, H.~S., {et~al.} 2011, \aap, 533, A119,
  \dodoi{10.1051/0004-6361/201117239}

\bibitem[{Fan {et~al.}(2016)Fan, Han, Nikutta, Drouart, \& Knudsen}]{Fan__2016}
Fan, L., Han, Y., Nikutta, R., Drouart, G., \& Knudsen, K.~K. 2016, The
  Astrophysical Journal, 823, 107, \dodoi{10.3847/0004-637x/823/2/107}

\bibitem[{Farrah {et~al.}(2017)Farrah, Petty, Connolly, Blain, Efstathiou,
  Lacy, Stern, Lake, Jarrett, Bridge, Eisenhardt, Benford, Jones, Tsai, Assef,
  Wu, \& Moustakas}]{Farrah_2017}
Farrah, D., Petty, S., Connolly, B., {et~al.} 2017, The Astrophysical Journal,
  844, 106, \dodoi{10.3847/1538-4357/aa78f2}

\bibitem[{Gonz{\'{a}}lez-Alfonso {et~al.}(2017)Gonz{\'{a}}lez-Alfonso, Fischer,
  Spoon, Stewart, Ashby, Veilleux, Smith, Sturm, Farrah, Falstad,
  Mel{\'{e}}ndez, Graci{\'{a}}-Carpio, Janssen, \&
  Lebouteiller}]{Gonz_lez_Alfonso_2017}
Gonz{\'{a}}lez-Alfonso, E., Fischer, J., Spoon, H. W.~W., {et~al.} 2017, The
  Astrophysical Journal, 836, 11, \dodoi{10.3847/1538-4357/836/1/11}

\bibitem[{Gowardhan {et~al.}(2018)Gowardhan, Spoon, Riechers,
  Gonz{\'{a}}lez-Alfonso, Farrah, Fischer, Darling, Fergulio, Afonso, \&
  Bizzocchi}]{Gowardhan_2018}
Gowardhan, A., Spoon, H., Riechers, D.~A., {et~al.} 2018, The Astrophysical
  Journal, 859, 35, \dodoi{10.3847/1538-4357/aabccc}

\bibitem[{{Greene} {et~al.}(2011){Greene}, {Zakamska}, {Ho}, \&
  {Barth}}]{greene_2011}
{Greene}, J.~E., {Zakamska}, N.~L., {Ho}, L.~C., \& {Barth}, A.~J. 2011, \apj,
  732, 9, \dodoi{10.1088/0004-637X/732/1/9}

\bibitem[{{Harrison} {et~al.}(2014){Harrison}, {Alexander}, {Mullaney}, \&
  {Swinbank}}]{harrison_2014}
{Harrison}, C.~M., {Alexander}, D.~M., {Mullaney}, J.~R., \& {Swinbank}, A.~M.
  2014, \mnras, 441, 3306, \dodoi{10.1093/mnras/stu515}

\bibitem[{{Heckman} {et~al.}(1990){Heckman}, {Armus}, \&
  {Miley}}]{Heckman_1990}
{Heckman}, T.~M., {Armus}, L., \& {Miley}, G.~K. 1990, \apjs, 74, 833,
  \dodoi{10.1086/191522}

\bibitem[{{Hopkins}(2012)}]{hopkins_2012}
{Hopkins}, P.~F. 2012, \mnras, 420, L8,
  \dodoi{10.1111/j.1745-3933.2011.01179.x}

\bibitem[{Jarrett {et~al.}(2011)Jarrett, Cohen, Masci, Wright, Stern, Benford,
  Blain, Carey, Cutri, Eisenhardt, Lonsdale, Mainzer, Marsh, Padgett, Petty,
  Ressler, Skrutskie, Stanford, Surace, Tsai, Wheelock, \& Yan}]{Jarrett_2011}
Jarrett, T.~H., Cohen, M., Masci, F., {et~al.} 2011, The Astrophysical Journal,
  735, 112, \dodoi{10.1088/0004-637x/735/2/112}

\bibitem[{{Jun} {et~al.}(2020){Jun}, {Assef}, {Bauer}, {Blain},
  {D{\'\i}az-Santos}, {Eisenhardt}, {Stern}, {Tsai}, {Wright}, \&
  {Wu}}]{Jun_2020}
{Jun}, H.~D., {Assef}, R.~J., {Bauer}, F.~E., {et~al.} 2020, \apj, 888, 110,
  \dodoi{10.3847/1538-4357/ab5e7b}

\bibitem[{Kang \& Woo(2018)}]{Kang_2018}
Kang, D., \& Woo, J.-H. 2018, The Astrophysical Journal, 864, 124,
  \dodoi{10.3847/1538-4357/aad561}

\bibitem[{{Kang} {et~al.}(2017){Kang}, {Woo}, \& {Bae}}]{kang_2017}
{Kang}, D., {Woo}, J.-H., \& {Bae}, H.-J. 2017, \apj, 845, 131,
  \dodoi{10.3847/1538-4357/aa80e8}

\bibitem[{Karouzos {et~al.}(2016{\natexlab{a}})Karouzos, Woo, \&
  Bae}]{Karouzos_2016}
Karouzos, M., Woo, J.-H., \& Bae, H.-J. 2016{\natexlab{a}}, The Astrophysical
  Journal, 833, 171, \dodoi{10.3847/1538-4357/833/2/171}

\bibitem[{Karouzos {et~al.}(2016{\natexlab{b}})Karouzos, Woo, \&
  Bae}]{Karouzos_2016a}
---. 2016{\natexlab{b}}, The Astrophysical Journal, 819, 148,
  \dodoi{10.3847/0004-637x/819/2/148}

\bibitem[{{Kewley} {et~al.}(2001){Kewley}, {Dopita}, {Sutherland}, {Heisler},
  \& {Trevena}}]{kewley_2001}
{Kewley}, L.~J., {Dopita}, M.~A., {Sutherland}, R.~S., {Heisler}, C.~A., \&
  {Trevena}, J. 2001, \apj, 556, 121, \dodoi{10.1086/321545}

\bibitem[{Kewley {et~al.}(2013)Kewley, Maier, Yabe, Ohta, Akiyama, Dopita, \&
  Yuan}]{Kewley_2013}
Kewley, L.~J., Maier, C., Yabe, K., {et~al.} 2013, The Astrophysical Journal,
  774, L10, \dodoi{10.1088/2041-8205/774/1/l10}

\bibitem[{Kim {et~al.}(2006)Kim, Ho, \& Im}]{Kim_2006}
Kim, M., Ho, L.~C., \& Im, M. 2006, The Astrophysical Journal, 642, 702,
  \dodoi{10.1086/501422}

\bibitem[{Kirkpatrick {et~al.}(2015)Kirkpatrick, Pope, Sajina, Roebuck, Yan,
  Armus, D{\'{\i}}az-Santos, \& Stierwalt}]{Kirkpatrick_2015}
Kirkpatrick, A., Pope, A., Sajina, A., {et~al.} 2015, The Astrophysical
  Journal, 814, 9, \dodoi{10.1088/0004-637x/814/1/9}

\bibitem[{Kormendy \& Ho(2013)}]{Kormendy_2013}
Kormendy, J., \& Ho, L.~C. 2013, Annual Review of Astronomy and Astrophysics,
  51, 511, \dodoi{10.1146/annurev-astro-082708-101811}

\bibitem[{{Lord}(1992)}]{Lord_1992}
{Lord}, S.~D. 1992, {A new software tool for computing Earth's atmospheric
  transmission of near- and far-infrared radiation}, NASA Technical Memorandum
  103957

\bibitem[{{Madau} \& {Dickinson}(2014)}]{Madau_2014}
{Madau}, P., \& {Dickinson}, M. 2014, \araa, 52, 415,
  \dodoi{10.1146/annurev-astro-081811-125615}

\bibitem[{Maiolino {et~al.}(2012)Maiolino, Gallerani, Neri, Cicone, Ferrara,
  Genzel, Lutz, Sturm, Tacconi, Walter, Feruglio, Fiore, \&
  Piconcelli}]{maiolino_2012}
Maiolino, R., Gallerani, S., Neri, R., {et~al.} 2012, Monthly Notices of the
  Royal Astronomical Society: Letters, 425, L66,
  \dodoi{10.1111/j.1745-3933.2012.01303.x}

\bibitem[{Melbourne {et~al.}(2012)Melbourne, Soifer, Desai, Pope, Armus, Dey,
  Bussmann, Jannuzi, \& Alberts}]{Melbourne_2012}
Melbourne, J., Soifer, B.~T., Desai, V., {et~al.} 2012, The Astronomical
  Journal, 143, 125, \dodoi{10.1088/0004-6256/143/5/125}

\bibitem[{Murphy {et~al.}(2011)Murphy, Chary, Dickinson, Pope, Frayer, \&
  Lin}]{Murphy_2011}
Murphy, E.~J., Chary, R.-R., Dickinson, M., {et~al.} 2011, The Astrophysical
  Journal, 732, 126, \dodoi{10.1088/0004-637x/732/2/126}

\bibitem[{{Nesvadba} {et~al.}(2011){Nesvadba}, {Polletta}, {Lehnert},
  {Bergeron}, {De Breuck}, {Lagache}, \& {Omont}}]{nesvadba_2011}
{Nesvadba}, N.~P.~H., {Polletta}, M., {Lehnert}, M.~D., {et~al.} 2011, \mnras,
  415, 2359, \dodoi{10.1111/j.1365-2966.2011.18862.x}

\bibitem[{{Osterbrock}(1989)}]{osterbrock}
{Osterbrock}, D.~E. 1989, {Astrophysics of gaseous nebulae and active galactic
  nuclei} ({University Science Books})

\bibitem[{{Osterbrock} \& {Ferland}(2006)}]{osterbrock_2006}
{Osterbrock}, D.~E., \& {Ferland}, G.~J. 2006, {Astrophysics of gaseous nebulae
  and active galactic nuclei} ({University Science Books})

\bibitem[{{Ramos Almeida} \& {Ricci}(2017)}]{almeida_2017}
{Ramos Almeida}, C., \& {Ricci}, C. 2017, Nature Astronomy, 1, 679,
  \dodoi{10.1038/s41550-017-0232-z}

\bibitem[{Rich {et~al.}(2010)Rich, Dopita, Kewley, \& Rupke}]{Rich_2010}
Rich, J.~A., Dopita, M.~A., Kewley, L.~J., \& Rupke, D. S.~N. 2010, The
  Astrophysical Journal, 721, 505, \dodoi{10.1088/0004-637x/721/1/505}

\bibitem[{Richards {et~al.}(2006)Richards, Lacy, Storrie-Lombardi, Hall,
  Gallagher, Hines, Fan, Papovich, Berk, Trammell, Schneider, Vestergaard,
  York, Jester, Anderson, Budavari, \& Szalay}]{Richards_2006}
Richards, G.~T., Lacy, M., Storrie-Lombardi, L.~J., {et~al.} 2006, The
  Astrophysical Journal Supplement Series, 166, 470, \dodoi{10.1086/506525}

\bibitem[{{Schmidt} {et~al.}(2018){Schmidt}, {Oio}, {Ferreiro}, {Vega}, \&
  {Weidmann}}]{Schmidt_2018}
{Schmidt}, E.~O., {Oio}, G.~A., {Ferreiro}, D., {Vega}, L., \& {Weidmann}, W.
  2018, \aap, 615, A13, \dodoi{10.1051/0004-6361/201731557}

\bibitem[{{Skrutskie} {et~al.}(2006){Skrutskie}, {Cutri}, {Stiening},
  {Weinberg}, {Schneider}, {Carpenter}, {Beichman}, {Capps}, {Chester},
  {Elias}, {Huchra}, {Liebert}, {Lonsdale}, {Monet}, {Price}, {Seitzer},
  {Jarrett}, {Kirkpatrick}, {Gizis}, {Howard}, {Evans}, {Fowler}, {Fullmer},
  {Hurt}, {Light}, {Kopan}, {Marsh}, {McCallon}, {Tam}, {Van Dyk}, \&
  {Wheelock}}]{2MASS}
{Skrutskie}, M.~F., {Cutri}, R.~M., {Stiening}, R., {et~al.} 2006, \aj, 131,
  1163, \dodoi{10.1086/498708}

\bibitem[{{Soto} {et~al.}(2012){Soto}, {Martin}, {Prescott}, \&
  {Armus}}]{Soto_2012}
{Soto}, K.~T., {Martin}, C.~L., {Prescott}, M.~K.~M., \& {Armus}, L. 2012,
  \apj, 757, 86, \dodoi{10.1088/0004-637X/757/1/86}

\bibitem[{Speagle {et~al.}(2014)Speagle, Steinhardt, Capak, \&
  Silverman}]{Speagle_2014}
Speagle, J.~S., Steinhardt, C.~L., Capak, P.~L., \& Silverman, J.~D. 2014, The
  Astrophysical Journal Supplement Series, 214, 15,
  \dodoi{10.1088/0067-0049/214/2/15}

\bibitem[{{Spoon} \& {Holt}(2009)}]{spoon_2009}
{Spoon}, H.~W.~W., \& {Holt}, J. 2009, \apjl, 702, L42,
  \dodoi{10.1088/0004-637X/702/1/L42}

\bibitem[{Tsai {et~al.}(2015)Tsai, Eisenhardt, Wu, Stern, Assef, Blain, Bridge,
  Benford, Cutri, Griffith, Jarrett, Lonsdale, Masci, Moustakas, Petty, Sayers,
  Stanford, Wright, Yan, Leisawitz, Liu, Mainzer, McLean, Padgett, Skrutskie,
  Gelino, Beichman, \& Juneau}]{Tsai_2015}
Tsai, C.-W., Eisenhardt, P. R.~M., Wu, J., {et~al.} 2015, The Astrophysical
  Journal, 805, 90, \dodoi{10.1088/0004-637x/805/2/90}

\bibitem[{Tsai {et~al.}(2018)Tsai, Eisenhardt, Jun, Wu, Assef, Blain,
  D{\'{\i}}az-Santos, Jones, Stern, Wright, \& Yeh}]{Tsai_2018}
Tsai, C.-W., Eisenhardt, P. R.~M., Jun, H.~D., {et~al.} 2018, The Astrophysical
  Journal, 868, 15, \dodoi{10.3847/1538-4357/aae698}

\bibitem[{Vayner {et~al.}(2017)Vayner, Wright, Murray, Armus, Larkin, \&
  Mieda}]{Vayner_2017}
Vayner, A., Wright, S.~A., Murray, N., {et~al.} 2017, The Astrophysical
  Journal, 851, 126, \dodoi{10.3847/1538-4357/aa9c42}

\bibitem[{{Veilleux} {et~al.}(2013){Veilleux}, {Mel{\'e}ndez}, {Sturm},
  {Gracia-Carpio}, {Fischer}, {Gonz{\'a}lez-Alfonso}, {Contursi}, {Lutz},
  {Poglitsch}, {Davies}, {Genzel}, {Tacconi}, {de Jong}, {Sternberg}, {Netzer},
  {Hailey-Dunsheath}, {Verma}, {Rupke}, {Maiolino}, {Teng}, \&
  {Polisensky}}]{Veilleux_2013}
{Veilleux}, S., {Mel{\'e}ndez}, M., {Sturm}, E., {et~al.} 2013, \apj, 776, 27,
  \dodoi{10.1088/0004-637X/776/1/27}

\bibitem[{Wilson {et~al.}(2004)Wilson, Henderson, Herter, Matthews, Skrutskie,
  Adams, Moon, Smith, Gautier, Ressler, Soifer, Lin, Howard, LaMarr, Stolberg,
  \& Zink}]{nires}
Wilson, J.~C., Henderson, C.~P., Herter, T.~L., {et~al.} 2004, in Ground-based
  Instrumentation for Astronomy, ed. A.~F.~M. Moorwood \& M.~Iye, Vol. 5492,
  International Society for Optics and Photonics (SPIE), 1295 -- 1305,
  \dodoi{10.1117/12.550925}

\bibitem[{{Woo} {et~al.}(2017){Woo}, {Son}, \& {Bae}}]{woo_2017}
{Woo}, J.-H., {Son}, D., \& {Bae}, H.-J. 2017, \apj, 839, 120,
  \dodoi{10.3847/1538-4357/aa6894}

\bibitem[{Wright {et~al.}(2010)Wright, Eisenhardt, Mainzer, Ressler, Cutri,
  Jarrett, Kirkpatrick, Padgett, McMillan, Skrutskie, Stanford, Cohen, Walker,
  Mather, Leisawitz, Gautier, McLean, Benford, Lonsdale, Blain, Mendez, Irace,
  Duval, Liu, Royer, Heinrichsen, Howard, Shannon, Kendall, Walsh, Larsen,
  Cardon, Schick, Schwalm, Abid, Fabinsky, Naes, \& Tsai}]{Wright_2010}
Wright, E.~L., Eisenhardt, P. R.~M., Mainzer, A.~K., {et~al.} 2010, The
  Astronomical Journal, 140, 1868, \dodoi{10.1088/0004-6256/140/6/1868}

\bibitem[{Wu {et~al.}(2012)Wu, Tsai, Sayers, Benford, Bridge, Blain,
  Eisenhardt, Stern, Petty, Assef, Bussmann, Comerford, Cutri, Evans, Griffith,
  Jarrett, Lake, Lonsdale, Rho, Stanford, Weiner, Wright, \& Yan}]{Wu_2012}
Wu, J., Tsai, C.-W., Sayers, J., {et~al.} 2012, The Astrophysical Journal, 756,
  96, \dodoi{10.1088/0004-637x/756/1/96}

\bibitem[{Wu {et~al.}(2018)Wu, Jun, Assef, Tsai, Wright, Eisenhardt, Blain,
  Stern, D{\'{\i}}az-Santos, Denney, Hayden, Perlmutter, Aldering, Boone, \&
  Fagrelius}]{Wu_2018}
Wu, J., Jun, H.~D., Assef, R.~J., {et~al.} 2018, The Astrophysical Journal,
  852, 96, \dodoi{10.3847/1538-4357/aa9ff3}

\bibitem[{Zakamska {et~al.}(2016)Zakamska, Hamann, Paris, Brandt, Greene,
  Strauss, Villforth, Wylezalek, Alexandroff, \& Ross}]{zakamska_2016}
Zakamska, N.~L., Hamann, F., Paris, I., {et~al.} 2016, Monthly Notices of the
  Royal Astronomical Society, 459, 3144, \dodoi{10.1093/mnras/stw718}

\end{thebibliography}
\bibliographystyle{aasjournal}

\appendix
\section*{[SII] and Other Line Measurements}
W0514-1217 is the only target for which observations allow a good constraint on the electron density through the \sii\ doublet ratio. The \sii$\lambda$6718/$\lambda$6733 ratio is $1.05\pm0.15$, corresponding to an electron density $n_e \approx 300\rm\ cm^{-3}$ \citep{osterbrock}.  This is slightly lower than the values reported for the two other Hot DOGs with a well-measured \sii\ ratio \citep{Jun_2020}, but is still typical of observed AGN \citep{osterbrock}.  In other targets, the line profile is too blended to clearly measure the ratio of the doublet components, though changes in the combined line center (e.g. W2238+2653 vs W2216+0723) suggest the densities may range below and above the regime for which \sii is a useful estimator.

\begin{figure*}
    \centering
    \noindent\includegraphics[width=39pc]{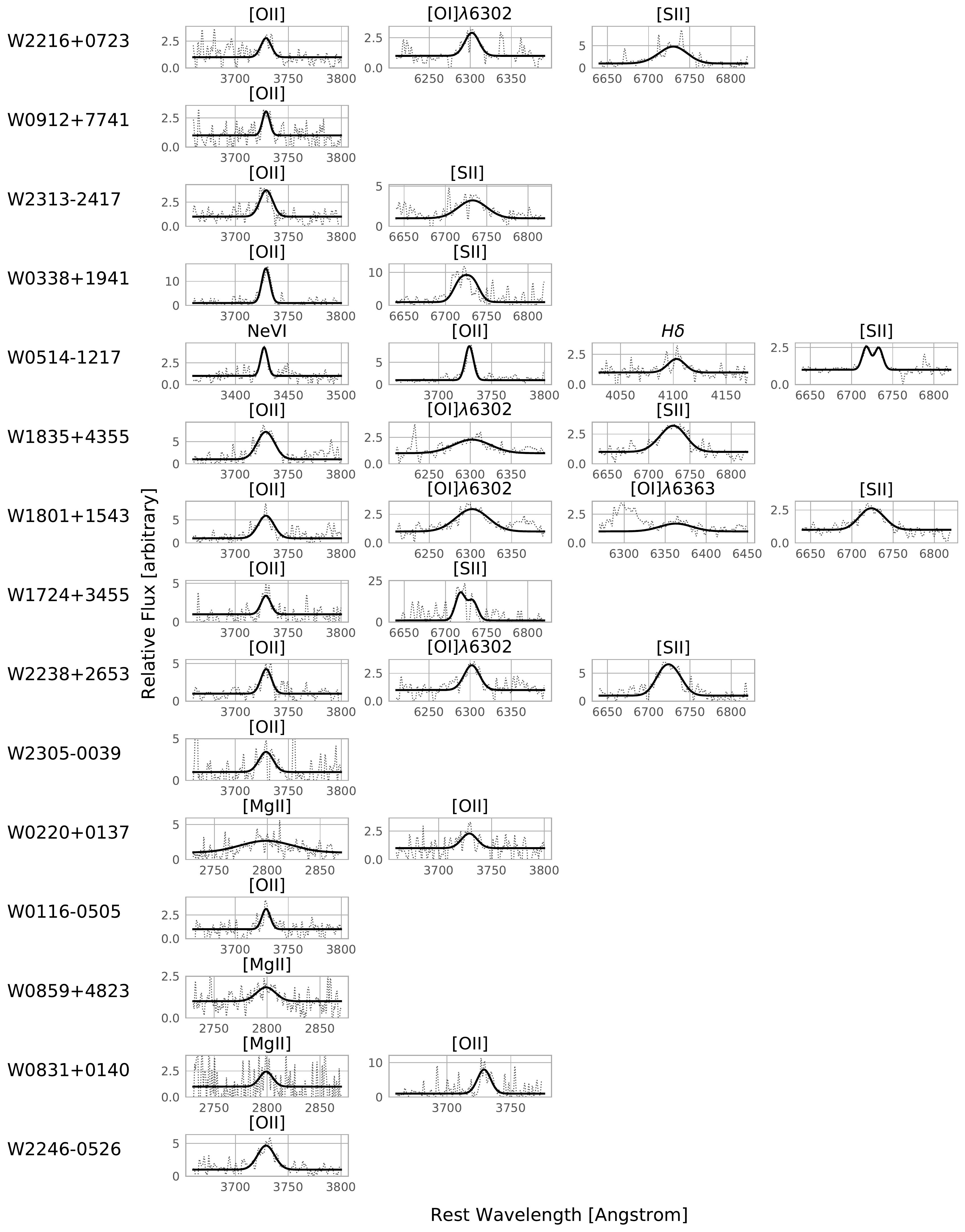}
    \caption{Other lines detected and fit.  All lines except \sii\ used a single Gaussian to fit redshift, intensity, and width.  \sii\ used the redshift determined from H$\alpha$ and fit intensity, width, and the ratio of the doublet. The blending of the \sii\ doublet limits its use as a density diagnostic, except in W0514-1217.}
    \label{lineplots3}
\end{figure*}

For several targets, profile fits were obtained for additional lines, mostly [\ion{O}{2}] or [\ion{O}{1}].  All of these lines were fit with a single Gaussian profile after continuum normalization, and generally suffer from poor signal-to-noise.  As a result, no attempt was made to fit outflows to any of these lines, though outflows have been previously claimed on [\ion{O}{2}] lines in \citet{Jun_2020}.  Residual tellurics are particularly problematic here as many of these lines were detected in the $J$ and $H$ bands where sky subtraction was poor, and telluric residuals or poor sky subtraction can leave fluctuations comparable in amplitude to the line itself. This is evident in the possible detection of $\rm H\delta$ in W0514-1217. Table \ref{otherlines} presents equivalent widths and FWHMs, and Figure \ref{lineplots3} shows the fit profiles.

\begin{deluxetable}{cccc}\centering
\tablewidth{0pt}
\tabletypesize{\scriptsize}
\tablecaption{Other Line Parameters}
\tablehead{ & & \colhead{FWHM [\kms]} & \colhead{EW [\r{A}]}}
\startdata
W0116-0505 & [OII]$\lambda3727$ & 800$\pm$200 & 90$\pm$30 \\  
W0220+0137 & [MgII]$\lambda 2799$ & 6200$\pm$600 & 430$\pm$50 \\
& [OII]$\lambda 3727$ & 1400$\pm$400 & 100$\pm$30 \\
W0338+1941 & [OII]$\lambda 3727$ & 730$\pm$30 & 430$\pm$30 \\
& [SII] & - & 600$\pm$60 \\
W0514-1217 & [NeVI]$\lambda 3427$ & 630$\pm$50 & 80$\pm$10 \\
& [OII]$\lambda3727$ & 740$\pm$30 & 240$\pm$10  \\
& HeI$\lambda 3889$ &  400$\pm$40 & 34$\pm$3 \\
& H$\delta$ & 1394$\pm$177 & 70$\pm$10 \\
& [SII] & - & 120$\pm$30 \\
W0831+0140 & MgII$\lambda 2799$ & 1600$\pm$800 & 110$\pm$70 \\
& [OII]$\lambda3727$ &  1000$\pm$100 & 450$\pm$80 \\
W0859+4823 & MgII$\lambda 2799$ &  600$\pm$200 & 50$\pm$20 \\
W0912+7741 & [OII]$\lambda3727$ &  650$\pm$80 & 50$\pm$10 \\
W1724+3355 & [OII]$\lambda3727$ &  1400$\pm$300 & 800$\pm$200 \\
& [SII] & - & 700$\pm$100 \\
W1801+1543 & [OII]$\lambda 3727$ & 1400$\pm$131 & 300$\pm$40 \\
& [OI]$\lambda 6302$ & 2200$\pm$300 & 320$\pm$50  \\
& [OI]$\lambda 6363$ & 2200$\pm$300 & 110$\pm$20 \\
& [SII] & - & 200$\pm$50 \\
W1835+4355 & [OII]$\lambda3727$ &  1500$\pm$100 & 420$\pm$40 \\
& [OI]$\lambda 6302$ &  2500$\pm$300 & 240$\pm$30 \\
& [SII] & - & 290$\pm$20 \\
W2216+0723 & [OII]$\lambda3727$ & 900$\pm$100 & 60$\pm$10  \\
& [OI]$\lambda 6302$ & 1000$\pm$10 & 110$\pm$20 \\
& [SII] & - & 890$\pm$70 \\
W2238+2653 & [OII]$\lambda3727$ &  1000$\pm$100 & 140$\pm$20 \\
& [OI]$\lambda 6302$ &   1100$\pm$100 & 180$\pm$30 \\
& [SII] & - & 590$\pm$50 \\
W2246-0526 & [OII]$\lambda3727$ &  1500$\pm$100 & 410$\pm$40 \\
W2313-2417 & [OII]$\lambda 3727$ & 1100$\pm$100 & 130$\pm$10 \\
& [SII] & - & 300$\pm$50 \\
\enddata
\tablecomments{[OIII]$\lambda 3727/\lambda 3730$ doublet is not well-resolved and treated as a single Gaussian. [SII] is the total equivalent width of the $\lambda\lambda$6718,6732 doublet.}
\label{otherlines}
\end{deluxetable}

\end{document}